\PassOptionsToPackage{table}{xcolor}                   %
\documentclass[a4paper,onecolumn,11pt]{quantumarticle}
\pdfoutput=1
\usepackage[utf8]{inputenc}
\usepackage[english]{babel}
\usepackage[T1]{fontenc}
\input{glyphtounicode}\pdfgentounicode=1                 %
\usepackage{amsmath,amssymb,amsthm}
\usepackage{graphicx}
\usepackage{booktabs}
\usepackage{array}
\usepackage{float}                                       %
\usepackage{needspace}                                   %
\newcolumntype{L}[1]{>{\raggedright\arraybackslash}p{#1}}   %
\usepackage{caption}
\usepackage[section]{placeins}                  %
\usepackage{enumitem}
\usepackage{hyperref}
\usepackage{xcolor}
\definecolor{ours}{HTML}{FDEBDD}                       %
\definecolor{stripe}{HTML}{F2F3F5}                     %
\graphicspath{{figures/}}
\setlist{itemsep=2pt,topsep=4pt}

\newcommand{\bra}[1]{\langle #1 |}
\newcommand{\ket}[1]{| #1 \rangle}

\begin{document}

\title{Seeing Why: Exact Evidence Maps for Quantum and Optical Image Classifiers}

\author{Teerawat Tongloy}
\orcid{0009-0001-0670-3124}
\email{teerawat.to@kmitl.ac.th}
\affiliation{School of Integrated Innovative Technology, King Mongkut's Institute of Technology Ladkrabang, Bangkok,
Thailand}
\author{Rangsan Jomtarak}
\orcid{0009-0009-5417-5571}
\email{rangsan\_jom@dusit.ac.th}
\thanks{corresponding author}
\affiliation{Faculty of Science and Technology, Suan Dusit University, Bangkok, Thailand}

\begin{abstract}
When an image classifier decides, we can usually only estimate why. For quantum and all-optical classifiers that
act linearly on amplitude-encoded pictures up to the measurement, with no trained classical network before or after
them, each class's Born probability above its no-information share, $P_g - b_g$, splits exactly into contributions of
the encoded amplitudes, shown as \emph{Born evidence maps} (BEM). The split itself follows from known
results on quadratic forms: with the transmitted light held fixed, it is what Integrated Gradients from a zero
baseline and the Shapley value give. What we add is a way to measure it on the hardware that classifies, from
class probabilities on a quantum chip or, in simulation, from camera frames on a lens bench, with no gradient circuit
and no light sent backwards; a $g\times g$ map takes $1 + 2g^2$ runs. Deleting the 5\,\% of pixels with the highest evidence drops
accuracy from 73\,\% to 9\,\% (one seed), as occlusion and gradient $\times$ input do, while plain gradients remove
much less. The classifiers share one architecture, a Fourier-plane mask followed by a brickwall circuit on qubits or
by phase plates on a lens bench, read by a fixed \emph{Born headroom} readout with no trained parameter; centring each
picture on its own mean lets dark features count as much as bright ones. In simulation, four phase plates at one
wavelength reach $72.3\,\%$ on Caltech-10 colour photographs. On IBM's ibm\_kingston, with the mask computed classically before approximate
loading, 300 test pictures were classified at $37.7\,\%$ against $10\,\%$ chance ($56.3\,\%$ in exact simulation),
and $2\times2$ maps of the filtered input of 30 test pictures drawn at random in advance, measured on the chip,
agreed with the exact ones ($r = 0.88$; standard estimator 0.82). The maps also revealed a shortcut in the data; as the trim that removed it
was chosen on test pictures, the Caltech-10 results are exploratory.
\end{abstract}

\maketitle

\section{Introduction}
\label{sec:intro}
Explanations of image classifiers are usually approximations. Gradients, class activation
maps~\cite{selvaraju2020gradcam} and their variants estimate which pixels mattered, but their maps need not add up to
the decision they explain. Methods whose maps do add up, such as Integrated Gradients~\cite{sundararajan2017ig} and
Shapley values~\cite{lundberg2017shap}, must in general be estimated numerically, by a path integral or by a sum over
feature coalitions.

Quantum and optical classifiers inherit the problem. Quantum classifiers range from convolution-like circuits,
introduced for quantum data~\cite{cong2019qcnn}, through amplitude-encoded classifiers~\cite{schuld2020circuit}, to
hybrids that place a pretrained classical network in front of the circuit~\cite{mari2020transfer}; in a hybrid, an
attribution concerns mainly the classical network. Hardware demonstrations, aimed at other questions, have typically
used a few input features (e.g.~\cite{havlicek2019feature}). Interpretability methods proposed for quantum
classifiers include gradient-based saliency through a hybrid network~\cite{lin2024qgradcam}, Integrated Gradients and
Shapley values of the input pixels, measured on a superconducting chip for $2\times2$ pictures~\cite{steinmuller2022xai},
Integrated Gradients for amplitude-encoded models~\cite{dibrita2025hattriq}, parameter-shift saliency on tiny
inputs~\cite{mehta2025qcxai}, and Shapley values over gates~\cite{heese2025shapley}; their opportunities and limits
are reviewed in~\cite{gilfuster2024explaining}. Optical diffractive
networks~\cite{lin2018d2nn} classify pictures with light alone, in simulation and in fabricated devices; the attributions we
found for them are gradient-based, computed by back-propagating gradients through the wave
propagation~\cite{huang2026gpam}. That map, the overlap of the forward field with the back-propagated detector field,
is close in form to ours. Ours adds up exactly to each class's probability above its share, can be measured from
camera intensities alone, and is tested on colour photographs; that work notes that the dark backgrounds of its
pictures may favour the foreground. Separately, the layers of trained diffractive networks have been interpreted as the
physical transforms they perform~\cite{wu2025physical}.

A classifier that is linear in its input amplitudes up to the final measurement, with no classical network before or
after it, does not need approximate explanations: the contribution of every input amplitude to every class's measured probability
can be written down in closed form. The explanation is then part of the model's design rather than an
approximation added afterwards, and accuracy is what the design costs. We build such classifiers for two kinds of
hardware, show that this
decomposition is exact, check its ranking of pixels in a deletion test, and measure it from device outputs ($2\times2$ blocks on
a quantum chip, per pixel on a simulated lens bench). Our contributions:
\begin{enumerate}
\item \emph{Born evidence maps} (BEM, Section~\ref{sec:maps}): an exact additive decomposition of each class's
  probability above its share over the encoded amplitudes, shown as spatial maps and, on hardware, estimated from
  finite measurements. The split itself is not new mathematics: for the quadratic numerator of each class's probability, with the light that
  arrives held fixed, it is what Integrated Gradients from a zero baseline and the Shapley value give, by known results
  on polynomials~\cite{lundstrom2023ig,deng2024taylor}, and the models of~\cite{dibrita2025hattriq} start from the
  same form. The exactness comes from the quadratic form, not from quantum mechanics; the Born rule (or, for light,
  intensity detection) is what makes that form physically realised, so the hardware produces it directly. What we
  add is the target, $P_g - b_g$ of a Born-rule classifier, to which the maps sum exactly, and a
  measurement of the maps from class probabilities or camera frames, without gradient
  circuits~\cite{dibrita2025hattriq} or the many shifted inputs of a numerical path integral or coalition
  sum~\cite{steinmuller2022xai} (next item). In a deletion test the maps rank pixels as well as
  occlusion and gradient $\times$ input, which with the transmission held fixed is twice the evidence, while plain
  gradients remove much less accuracy (seed 0, 300 pictures; Section~\ref{sec:deletion}). The test checks the
  ranking; what the maps add is the exact sum and its measurement.
\item \emph{Maps measured on hardware} (Section~\ref{sec:measure}). A known identity, the polarization of a
  quadratic form, of the kind used to replace Hadamard tests by direct measurements~\cite{mitarai2019direct}, applied with complementary blocks of the picture as its two parts, gives
  each block's exact share from three measured class probabilities, with no ancilla, gradient circuit, phase
  measurement or backward light. On IBM's ibm\_kingston, with the mask computed
  before loading, $2\times2$ maps of 30 test pictures drawn at random before any map circuit was fitted or run
  correlate with the exact maps at 0.88 with a sign-flip estimator (0.82 with the standard one; both fixed in advance), as well for the pictures the chip classified wrongly as for those it classified correctly
  (Section~\ref{sec:hardware}); on a simulated lens bench, per-pixel maps
  follow from camera intensities alone (Section~\ref{sec:bench}).
\item \emph{The mean-centred encoding} (Section~\ref{sec:meanplane}): an amplitude encoding that gives dark features the same
  weight as bright ones. It is the largest gain among the parts before the readout, and it moves the input light, and
  with it the evidence, onto the object, measured against human-drawn outlines (Section~\ref{sec:object}).
\item \emph{One architecture for qubits and lenses} (Sections~\ref{sec:mask}--\ref{sec:circuits}): a Fourier-plane mask
  (a quantum Fourier transform on a chip, a 4f system on a bench) followed by a brickwall circuit or by phase plates,
  trained on four datasets (five settings) from handwritten digits to colour photographs (Section~\ref{sec:standard}). The phase plates
  were checked end to end in an independent optics simulator, and we report the tolerances a bench must meet
  (Section~\ref{sec:builders}).
\item \emph{Born headroom} (Section~\ref{sec:bhr}): a readout with no trained parameter that scores each class by how
  far its measured probability rises above its no-information share. On the optical model it reaches 72.3\,\%, against
  48.6\,\% for the plain sums of the same outcomes and 73.8\,\% for a softmax with a learned temperature: BHR comes
  close to a learned temperature without training one, and is not more accurate (Section~\ref{sec:ablation}).
\item \emph{Explanations in use} (Section~\ref{sec:shortcut}): the maps exposed a shortcut, a dark frame on one class's
  pictures; we removed it and confirmed the removal with the same maps.
\end{enumerate}
In short, this is an optical image classifier whose class probabilities decompose exactly into contributions of the
encoded amplitudes (designed
and checked in simulation), and a quantum classifier with no trained classical network besides its linear Fourier-plane mask (computed off the
chip for the hardware runs) whose $2\times2$ explanations
were measured on a quantum computer. On
hardware, two steps run classically before the chip: for the masked model, the Fourier-plane mask, which a bench does
with lenses, and the fitting of a short circuit that loads each picture (Section~\ref{sec:hardware}).
\begin{figure}[!htb]
\centering
\includegraphics[width=\linewidth]{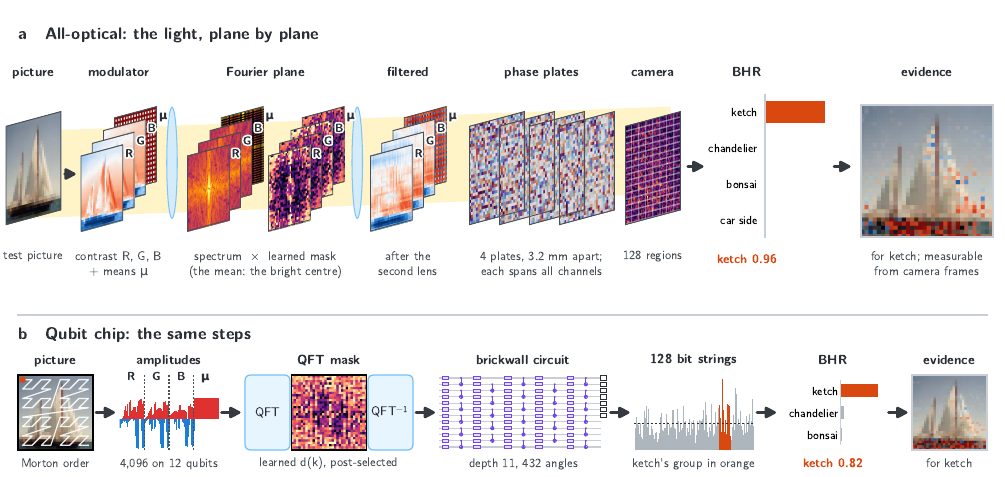}
\caption{Overview, drawn from the trained models' own data for a test picture of a ketch. (a) The all-optical
classifier along its optical axis. The modulator holds the encoded field: three colour planes, each with the picture's
own mean removed, and a plane of the colour means. A first lens takes each channel to its Fourier plane (its mean is
the bright centre), where the learned mask multiplies it; a second lens forms the filtered field. Four phase plates
(phase shown as colour) follow, and the Born headroom readout turns the light on the camera's 128 regions into class
probabilities. The Born evidence map (BEM) for the answer, which a bench can measure from camera frames, lies on the dark hull. (b) The qubit classifier takes the same steps
with a quantum Fourier transform, a brickwall circuit and 128 measured bit strings; on the chip of
Section~\ref{sec:hardware} the mask was applied before loading. The same picture through the trained models, plane by plane: Appendix~\ref{app:inside}.}
\label{fig:overview}
\end{figure}

\section{Method}
\label{sec:method}
Figure~\ref{fig:overview} shows the pipeline. We describe its parts in the order the light passes through them: the
encoding, the Fourier-plane mask, the trained circuit or plates, and the readout. The evidence maps then follow from
linearity, and we close with how they are measured on hardware.

\subsection{Encoding that respects the dark: the mean-centred encoding}
\label{sec:meanplane}
Amplitude encoding of raw intensities is positive everywhere. The light then follows the bright parts of a picture: a
bright sky outweighs a dark hull, and dark features carry little signal however informative they are. We remove each
picture's own mean from each colour plane. A dark detail then becomes a large negative amplitude, with the same energy
as an equally strong bright one. The three means move to the encoding's otherwise empty fourth plane, the \emph{means
plane}: a $2\times 2$ tile repeated over the grid holds the red, green and blue means and one dark cell, scaled so that
this plane carries the fraction $f = 0.5$ of the picture's energy ($f$ was set before any run; row D trained
again with $f = 0.25$ and $0.75$ gives $71.8 \pm 2.0$ and $72.6 \pm 0.2\,\%$ against $72.3 \pm 0.7\,\%$, three seeds each). With $c$ the three colour planes of the picture,
each with its own mean $\mu_R$, $\mu_G$ or $\mu_B$ removed, and $m$ the means plane (the tile $\mu_R, \mu_G, \mu_B,
0$ repeated), the encoded amplitudes are
\begin{equation}
  a = \Big(\sqrt{1-f}\,\frac{c}{\|c\|},\; \sqrt{f}\,\frac{m}{\|m\|}\Big),
  \label{eq:meanplane}
\end{equation}
in the qubit or field order of Section~\ref{sec:circuits}. Only the ratios of the three means and the relative sign of $c$ and $m$
survive this scaling, not their overall level. A picture of one flat colour has $c = 0$ and is not encoded; none occurs in
the data. The colour planes thus keep no mean (DC) at all, yet the circuit still receives the colour balance and a
phase reference; without that reference the sign of a deviation
would be lost, since $P(\psi) = P(-\psi)$. Optically, the means plane is the zero frequency, carried as a separate
reference tile; like the other planes, it passes the Fourier-plane mask with its own gain. In our setting the picture
reaches the bench as a digital image, so the three means are computed while the modulator is written. The centred
planes are signed, so the picture modulator writes an amplitude together with a phase of 0 or $\pi$; such complex
modulation is established, for example by pairing pixels on a single liquid-crystal modulator~\cite{arrizon2003double}
or by cascading two modulators. Only a sign is needed, which is simpler than full complex modulation; a double-pixel
scheme pays for it with at least two modulator pixels per picture pixel and a spatial filter, which the 4f system's
Fourier plane can provide. An optical stage
could instead separate each colour's zero frequency in a Fourier plane and expand it into the reference tile;
reproducing Eq.~\eqref{eq:meanplane} would additionally need an input-dependent balance of the contrast and reference
energies, which our implementation computes digitally before the modulator is written. Grey pictures (one plane on eleven qubits) carry their single mean evenly over the second plane, their means plane,
which is then a constant reference and carries no information about the picture's brightness. No dataset
statistic is used: every picture is centred on its own mean.

\subsection{Fourier-plane mask}
\label{sec:mask}
Before the main circuit, each colour plane passes a learned filter in its spatial-frequency plane, as in optical
classifiers with a learned Fourier-plane element~\cite{chang2018hybrid,yan2019fourier}: the plane is taken
to its two-dimensional Fourier transform, multiplied by a learned complex mask $d(k)$ (one mask shared by the planes,
with a learned gain per plane), and brought back. On a lens bench each plane passes its own 4f system, a $2\times2$
array of lens pairs with a spatial light modulator in their shared focal plane, so that the planes' filtered light does
not mix; on qubits it is a quantum Fourier transform on the row qubits and one on the
column qubits, with the mask applied by post-selected ancillas. The mask's amplitude is even in $k$ and its phase odd,
so the filter is a convolution with a real kernel; it is shift-equivariant and favours no place in the picture. The
qubit version, the \emph{circular mask}, convolves circularly. The lens version, the \emph{lens mask}, zero-pads the
picture to twice its size and keeps the original frame, so that light leaving the frame is lost, as on a bench. Being
lossy, the mask makes all probabilities conditional (Section~\ref{sec:maps}); its transmission is reported in
Table~\ref{tab:ablation}. The mask has an amplitude and a phase, so its modulator, like the picture's, needs complex
modulation, for example amplitude encoded onto a phase-only modulator~\cite{davis1999encoding}.

\subsection{Main circuits: qubit brickwall and phase plates}
\label{sec:circuits}
\emph{Qubits.} A $32\times32$ colour picture with its means plane fills $2^{12}$ amplitudes on twelve qubits, ordered
by plane (the two leading address bits) and within a plane along a Morton (Z-order) curve: the remaining ten bits
interleave row and column bits from the most significant down, which organises the pixels into nested, spatially coherent blocks. This orders the basis-state addresses; it does not make gates on neighbouring wires act on neighbouring pixels. The trained circuit is a
brickwall of depth $L$: $L+1$ layers of single-qubit rotations $R_X R_Y R_Z$ on every qubit, with nearest-neighbour
CNOTs on alternating pairs between layers. The first seven qubits are measured; their $2^7 = 128$ outcomes form the BHR
groups. Depth 11 has 432 parameters (1,460 with the mask).

\emph{Phase plates.} For an all-optical classifier the brickwall is replaced by $L$ phase plates. The four planes are
tiled $2\times2$ on one field of $64\times64$ pixels. Between plates of $32\times32$ phase pixels, light propagates by
Fresnel diffraction, with transfer function $\exp[-i\pi z(u^2+v^2)]$ on a grid padded twice; $z = \lambda d/\Delta x^2 =
32$ corresponds to 3.2\,mm at 633\,nm with 8\,\textmu m field samples (a plate pixel spans $2\times2$ of them, 16\,\textmu m). A camera at the end is cut into $16\times 8 = 128$
regions, the BHR outcomes. Everything is linear in the field, so the evidence maps carry over unchanged: $A^\dagger$ is
the phase-conjugated return path. The plain model samples each pixel at one point; a \emph{pixel-aware} variant models
every picture and plate pixel as a square, sampled $2\times2$ times, so that training sees the diffraction of the
pixels themselves. Four plates with the mask have 5,124 parameters.

\subsection{Born headroom readout}
\label{sec:bhr}
A classifier ends in a measurement with $N$ outcomes and Born probabilities $P_b$, $b = 1,\dots,N$: on a qubit chip the
$2^{n_\mathrm{out}}$ bit strings of its $n_\mathrm{out}$ measured qubits, on a lens bench the light falling on $N$
camera regions. The outcomes are split into $G$ contiguous groups $S_g$, one per class, and the class mass is $P_g =
\sum_{b\in S_g} P_b$. A maximally mixed input, which carries no information about the picture, gives class $g$ the mass
$P_g = b_g = |S_g|/N$ under any unitary circuit. We call $b_g$ the class's no-information share; where losses or
non-unital noise move the mixed state's group masses, we keep $b_g$ as a fixed anchor. The \emph{Born headroom} of class
$g$ is how far its measured probability rises above that share, in units of the room above it:
\begin{equation}
  u_g = \frac{P_g - b_g}{1 - b_g}, \qquad u_g \in \Big[-\frac{b_g}{1-b_g},\, 1\Big].
  \label{eq:headroom}
\end{equation}
Every class has the same ceiling $u_g = 1$, whatever the size of its group. The Born headroom readout (BHR) turns the
headrooms into class probabilities with no trained parameter,
\begin{equation}
  \hat p = \operatorname{softmax}\!\big(\kappa\, \mathrm{LeakyReLU}_\alpha(u)\big), \qquad
  \kappa = \frac{\ln\!\big[(1-\varepsilon)(G-1)/\varepsilon\big]}{\varphi}.
  \label{eq:bhr}
\end{equation}
Here $\alpha$ is the LeakyReLU's slope below zero, and $\varepsilon = \alpha = 0.01$ are fixed. The one free choice is
$\varphi \in (0,1)$, which sets $\kappa$ through a reference configuration of the headrooms: one equal to $\varphi$
and all others zero, at which the softmax calls that class with probability $1 - \varepsilon$. This reference fixes
the softmax's scale; it is not a jointly attainable set of class masses, since the anchors already sum to one. In
every run we use $\varphi = 0.0556 = (1.5 - 1)/(G - 1)$, for equal groups the headroom of a
class that holds 1.5 times its share (our 128 outcomes form ten groups of 12 or 13, nearly equal); it was fixed before the experiments reported here and not tuned per dataset.
Training minimises the cross-entropy of $\hat p$. All trainable parameters belong to the quantum or optical
transformation: on qubits the mask and the circuit's angles, on a bench the mask and the plates' phases. The readout
contains no trainable parameter; it is a fixed rule, through which the gradient still reaches those parameters, and the
classical optimiser that updates them is no layer on the path from the picture to the answer. Because this scale is set by a headroom
the circuit can reach rather than by a probability of one, the cross-entropy's gradient becomes small once the true
class leads the competing classes by a sufficient headroom.

\subsection{Born evidence maps}
\label{sec:maps}
Let the input be an amplitude-encoded picture $a$ (real, $\|a\| = 1$). A linear map $A$ takes it to the measurement
plane: a unitary circuit, or a mask followed by a circuit or by free-space propagation. $\Pi_g$ projects onto the
outcomes of class $g$. A lossy element such as a mask or an aperture makes the measured probabilities conditional on
the light that arrives, $P_g = \bra{a} A^\dagger \Pi_g A \ket{a} / Z$ with $Z = \bra{a} A^\dagger \Pi A \ket{a}$, $\Pi = \sum_g \Pi_g$
the light that reaches the measured outcomes (on a bench, the detector regions; $\Pi = 1$ for a circuit measured in
full). Writing
$M_g = A^\dagger (\Pi_g - b_g\Pi) A$, the \emph{evidence} of amplitude $x$ (one pixel of one plane) for class $g$, displayed as a Born evidence map (BEM), is
\begin{equation}
  e_g(x) = \frac{a(x)\,\mathrm{Re}\big[M_g\, a\big](x)}{Z}, \qquad \sum_x e_g(x) = P_g - b_g .
  \label{eq:evidence}
\end{equation}
The sum is exact, not an approximation: the evidence of the amplitudes adds up to the class's probability above its
share. \emph{Exact} means three things in this paper: the decomposition is an algebraic identity of the model; our
simulator computes it to numerical precision; on hardware it is estimated from a finite number of shots or camera
frames, with the errors reported there. An exact sum does not make this the only possible decomposition, nor a
proof of cause in the scene: it is the one that Integrated Gradients and the Shapley value give for this form, and
the deletion tests (Section~\ref{sec:deletion}) are a separate check of what it ranks. Positive evidence pushes the class above its share; negative evidence pulls it below. The decomposition is
exact for the encoded amplitudes. The encoding subtracts each picture's channel means, fills the means plane and
normalises the result (Section~\ref{sec:meanplane}), so a change of one raw pixel moves the encoded amplitudes all over
the picture; the maps are therefore not attributions to raw pixels through the encoder. A pixel map, as shown in our figures, sums the encoded evidence of the three colour planes of each
pixel, a spatial evidence map. The means plane belongs to no pixel, so its evidence is left
out and a pixel map adds up to $P_g - b_g$ minus that plane's part; the hardware blocks of Section~\ref{sec:measure}
include all four planes. The maps are additive over any set of pixels, so block maps are sums of the pixel map.
The part left out is not always small. Over the 300 test pictures and three seeds, the means plane carries a median
3\,\% of $P_g - b_g$ for the phase plates (quartiles $-22$ and $+80\,\%$) and 31\,\% for the qubit circuit (quartiles
$+5$ and $+74\,\%$), and more than $|P_g - b_g|$ itself in 26\,\% and 20\,\% of the pictures: a pixel map shows where
the evidence of the picture's contrast lies, not the whole of $P_g - b_g$.

Two further maps come from the same quantities: the \emph{attention} $a(x)\,\mathrm{Re}[A^\dagger A\, a](x)/Z$, each
pixel's share of the light that reaches the measurement plane, and the \emph{input intensity} $a(x)^2$ itself. The
attention adds up to one but is signed: where interference removes light, a pixel's share is negative. In our models
its negative part is small (on average 2--4\,\% of the sum of its absolute values, at most 10\,\%), and the figures show it on a
scale on which negative values are darkest. All
three need one forward and one adjoint pass, which a simulator computes exactly and which hardware can measure
(Section~\ref{sec:measure}). Because the numerator $\bra{a}M_g\ket{a}$ is a quadratic form, the evidence coincides
with Integrated Gradients from a zero baseline~\cite{sundararajan2017ig} of $f(v) = v^\top M_g v / Z(a)$, the
numerator with the denominator held at its value for the picture explained: the gradient along the path $ta$ is
$2t\,\mathrm{Re}[M_ga]/Z(a)$, and its integral gives Eq.~\eqref{eq:evidence}. It also coincides with the Shapley value of the game on
$f$ that switches amplitudes off and splits each pair term $a(x)[M_g]_{xy}a(y)$ equally between $x$ and $y$. Both
follow from known results on polynomials: from a zero baseline, Integrated Gradients share a monomial among its
variables in proportion to their exponents and the Shapley value shares it equally~\cite{lundstrom2023ig,deng2024taylor};
for a quadratic form the two coincide. The exactness is therefore a property of quadratic forms, classical
mathematics. What the Born rule contributes is physics: a measurement of the class's outcomes produces this quadratic
form of the input amplitudes directly, $P_g - b_g$, so the form exists in the hardware and not only in a model of
it; on a lens bench the square-law detection of intensities does the same for classical light. That is why the
decomposition can be measured, block by block, on the device that classifies (Section~\ref{sec:measure}).
Neither statement holds for $P_g$ itself as a function of the input, which does not change when the input is scaled,
so that a path from zero tells nothing about it; nor do the maps attribute the readout's output $\hat p$, which
follows through a LeakyReLU and a softmax. They decompose $P_g - b_g$ of the picture explained. Wherever this paper says that the maps add up to a class's
probability, it means this measured class mass $P_g$, never the readout's $\hat p_g$. The models
of~\cite{dibrita2025hattriq} start from the same quadratic form, $\bra{x}U^\dagger O U\ket{x}$, but read it
through a $\tanh$ and encode the pixels with an extra overflow amplitude; their Integrated Gradients, from a blank
picture, are computed numerically from gradients measured with a Hadamard test. Here the attribution targets the quadratic
numerator with the transmission held fixed, the decomposition is closed-form, and it is measured from class probabilities alone
(Section~\ref{sec:measure}).

\subsection{Measuring the maps on hardware}
\label{sec:measure}
\emph{On a qubit chip}, the evidence of a block $S$ of pixels follows from three class probabilities by the
standard polarization identity of a quadratic form, $2\,\mathrm{Re}\,\bra{u}M\ket{v} = \bra{u+v}M\ket{u+v} - \bra{u}M\ket{u} - \bra{v}M\ket{v}$, applied to $u = \psi_S$, $v = \psi_{\bar S}$:
\begin{equation}
  E_S = \tfrac12\Big(\bra{\psi} M_g \ket{\psi} + \bra{\psi_S} M_g \ket{\psi_S} - \bra{\psi_{\bar S}} M_g \ket{\psi_{\bar S}}\Big),
  \label{eq:polar}
\end{equation}
where $E_S$ is the evidence of block $S$ for class $g$, $\psi$ is the chip's input, $\psi_S$ its part inside the
block and $\psi_{\bar S}$ the rest, and $M_g$ is now the circuit's own operator. The chip's input is the renormalised
filtered state, so $Z = 1$ and no division appears; on a bench (Eq.~\eqref{eq:bench}) the division by the transmitted
light remains. Reading cross terms from expectation values of prepared superpositions instead of a
Hadamard test is established in near-term quantum algorithms~\cite{mitarai2019direct,cortes2022krylov}; the same
identity, read from intensities, underlies in-situ gradient measurement in photonic networks~\cite{hughes2018trim,
pai2023insitu}. Here the two parts are complementary blocks of the input itself: their sum is the state already
prepared for the prediction, so no ancilla, reference state or relative phase is needed, and the block terms add up to
$P_g - b_g$, which checks the measurement. Equivalently, $E_S$ is the mean of the block's occlusion drop,
$f(\psi) - f(\psi_{\bar S})$, and its value alone, $f(\psi_S)$, with $f(v) = \bra{v}M_g\ket{v}$; for this quadratic
form that is the block's exact Shapley value, the closed form of the Shapley value of a game with only pairwise
interactions (a 2-additive game~\cite{grabisch1997kadditive}). Occluding each block alone would take only $1 + g^2$
runs, but its drops do not add up to $P_g - b_g$; the extra $g^2$ runs buy the exact, additive share. Each term $\bra{v}M_g\ket{v} = \|v\|^2 (P_g(v) - b_g)$ is a measured class
probability of a prepared state. In our hardware runs $\psi$ is the masked picture, so these maps attribute to blocks
of the filtered input. A $g\times g$ map needs $1 + 2g^2$ circuits; the identity is exact, and the measured blocks add
up to $P_g - b_g$ in expectation. The identity's other form, $4\,\mathrm{Re}\,\bra{u}M\ket{v} = \bra{u+v}M\ket{u+v} -
\bra{u-v}M\ket{u-v}$, gives a second estimator,
\begin{equation}
  E_S = f(\psi_S) + \tfrac14\big(f(\psi) - f(\psi_S - \psi_{\bar S})\big),
  \label{eq:flip}
\end{equation}
where $\psi_S - \psi_{\bar S}$ is the picture with the rest of it negated, again of unit norm. It also takes
$1 + 2g^2$ circuits, but weights the measured terms less; on the chip it gave the more accurate maps
(Section~\ref{sec:hardware}). With $s(v) = P_g(v)(1-P_g(v))/N$ the shot variance of a circuit's class probability
($N$ shots) and $w = \|\psi_S\|^2$, a block's variance is $[s(\psi) + w^2 s(\psi_S) + (1-w)^2 s(\psi_{\bar S})]/4$
by Eq.~\eqref{eq:polar} and $w^2 s(\psi_S) + [s(\psi) + s(\psi_S - \psi_{\bar S})]/16$ by Eq.~\eqref{eq:flip}.
With the exact probabilities of the circuits sent for the 30 random pictures (for $4\times4$, the four pictures prepared) and 4,000 shots, a $2\times2$ block's
standard deviation is 0.0021 by the sign flip and 0.0032 by Eq.~\eqref{eq:polar}, so the standard estimator needs
2.3 times the shots for the same error; the sign flip's implied mean absolute error, 0.0017, is the shot floor that
simulation gives. A Hadamard test of the same block needs fewer shots than the sign flip (0.6 times at $2\times2$, 0.2 times at
$4\times4$, as it reads each block in one of $g^2$ circuits), but an ancilla and controlled preparations of both
$\psi_S$ and $\psi$, on hardware where, for the masked model D$'$, preparing the picture alone already costs the most accuracy
(Section~\ref{sec:hardware}); the polarization route saves circuit depth, not shots. \emph{On a lens bench} the same identity needs only camera intensities: the input
modulator shows, in turn, the whole picture, only its part inside the block and only its part outside it (the same
pixels in all four tiles), and each time the camera records the light $I_r$ on every detector region $r$. With weights
$W_r = 1 - b_g$ on the class's regions and $-b_g$ elsewhere, $\bra{v} M \ket{v} = \sum_r W_r I_r(v)$ for any input
$v$, so
\begin{equation}
  E_S = \frac{W\!\cdot\! I(a) + W\!\cdot\! I(a_S) - W\!\cdot\! I(a_{\bar S})}{2 \sum_r I_r(a)} .
  \label{eq:bench}
\end{equation}
No phase is measured and no light is sent backwards: both fields are forward inputs on the modulator, unlike the
adjoint field of in-situ backpropagation~\cite{hughes2018trim,pai2023insitu}, and the rest of the picture plays the
part of the static reference used in transmission-matrix measurement~\cite{popoff2010tm,yoon2015tm}. A per-pixel map ($32\times32$ blocks) takes 2,049 camera frames,
at least 20 seconds at 100 frames per second (more with the modulator's settling time).

\section{Results}
\label{sec:results}
We first describe the data (Section~\ref{sec:data}), then what each part adds (Section~\ref{sec:ablation}) and how the
model fares on four datasets (Section~\ref{sec:standard}). We then test the maps: whether they point at what the model
uses (Section~\ref{sec:deletion}), whether they sit on the object (Section~\ref{sec:object}), and whether they find a
flaw in the data (Section~\ref{sec:shortcut}). Finally we measure them on a quantum computer and on a simulated lens
bench (Sections~\ref{sec:hardware}--\ref{sec:bench}) and give the tolerances a bench must meet
(Section~\ref{sec:builders}).

\subsection{Data and protocol}
\label{sec:data}
The main set is Caltech-101~\cite{caltech101} (CC BY 4.0), its ten largest object classes (airplane, motorbike, face,
watch, leopard, bonsai, car side, ketch, chandelier, hawksbill), called Caltech-10 below: all their pictures, 2,644 for
training, 100 held out for validation and 300 for testing (30 per class), with class-balanced batches. Each picture
loses 4\,\% of every side before a square centre crop to $32\times32$ colour pixels. The trim removes a black frame that
one class carried, a shortcut the maps themselves revealed (Section~\ref{sec:shortcut}); the crops without it are the
\emph{untrimmed} set. The trim was chosen after the shortcut was seen in maps of test pictures, so every result on
the trimmed Caltech-10 below is exploratory (Section~\ref{sec:limits}). Training uses random square crops of side $U(0.8, 1)$ of the raw picture, resized back, so no
artificial edge is created. Every pixel and colour of the $32\times32$ picture enters the model as an amplitude: there
is no PCA, pooling or feature extraction, learned or hand-made. For CIFAR-10 this is the dataset's full resolution and
for MNIST and Fashion-MNIST the full $28\times28$ picture, padded to $32\times32$; the Caltech-101 originals are larger
and are reduced to $32\times32$ by the crop above.

On Caltech-10 we train for 50 epochs with three seeds and report mean $\pm$ standard deviation. With 300 test pictures,
one point of accuracy is three pictures and a single run's 95\,\% interval is about $\pm 5.5$ points, so we report
ROC-AUC (macro, one-vs-rest) beside every accuracy. The standard sets are MNIST and Fashion-MNIST (grey, 58k training,
2k validation, full test sets) and CIFAR-10 in grey and in colour (48k / 2k), 20 epochs. All runs use BHR with $\varphi
= 0.0556$ and the cross-entropy of $\hat p$, Adam~\cite{kingma2015adam} with learning rate 0.01, batches of 64, and a
one-epoch linear warm-up followed by a cosine decay to zero. Rotation angles start from $\mathcal N(0, 0.1^2)$ and plate
phases from $\mathcal N(0, 1)$. The mask's amplitude, $\cos\theta_\mathrm{plane}\cos\theta_k$, starts at about one half,
with both angles at $\pi/4$ plus the same noise; its phases start near zero. The weights of the epoch with the best
validation accuracy are kept and evaluated once on the test set.

\begin{figure}[!htb]
\centering
\includegraphics[width=\linewidth]{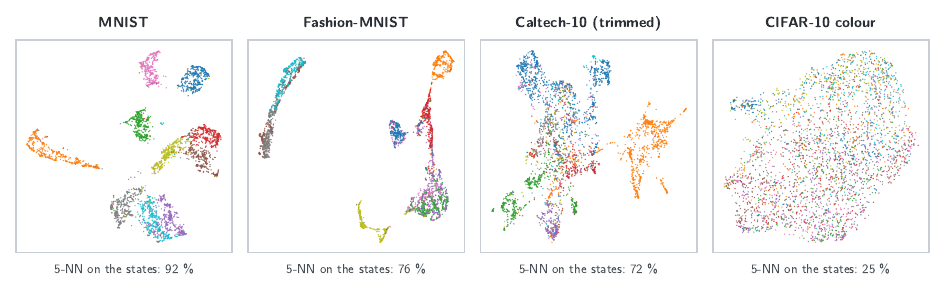}
\caption{How hard each dataset is: UMAP~\cite{mcinnes2018umap} of the plainly amplitude-encoded states (cosine distance; 3,000 pictures drawn at random from each test set, and for Caltech-10
from all its 3,044 pictures), coloured by class, with the 5-nearest-neighbour accuracy in the state space itself (1,500
pictures against the other 1,500).}
\label{fig:umap}
\end{figure}

Our simulator computes states, probabilities and gradients in PyTorch~\cite{paszke2019pytorch} and agrees with
PennyLane~\cite{bergholm2018pennylane} (to $10^{-10}$) and with Qiskit's~\cite{javadi2024qiskit} statevector (to
$10^{-16}$); the phase-plate propagation agrees with LightPipes'~\cite{lightpipes} Fresnel propagator to
$2\times10^{-10}$. Figure~\ref{fig:umap} shows how hard the four sets are for a classifier of the raw amplitude states (UMAP with cosine distance, which
orders states like fidelity for non-negative amplitudes):
MNIST separates almost by itself, CIFAR-10 hardly at all. In the charts, orange marks our main model or result and blue
or grey what it is compared with; in the maps, red and blue mean evidence for and against a class.

\subsection{What each part adds}
\label{sec:ablation}
\begin{table}[t]
\centering\small
\setlength{\tabcolsep}{2.5pt}
\begin{tabular}{llrccc}
\toprule
row & model & params & accuracy (\%) & ROC-AUC & transm. \\
\midrule
A  & phase plates, plain amplitudes            & 4,096 & $61.4 \pm 0.8$ & $0.892 \pm 0.005$ & 0.419 \\
B  & + mean-centred encoding                & 4,096 & $68.2 \pm 2.2$ & $0.939 \pm 0.005$ & 0.405 \\
C  & + lens mask                               & 4,614 & $68.9 \pm 1.1$ & $0.940 \pm 0.008$ & 0.044 \\
\rowcolor{ours}D  & + phase in the mask                       & 5,124 & $\mathbf{72.3 \pm 0.7}$ & $0.947 \pm 0.008$ & 0.042 \\
E  & D, pixel-aware plates                     & 5,124 & $71.6 \pm 1.7$ & $0.947 \pm 0.010$ & 0.020 \\
F  & E + light-keeping loss                    & 5,124 & $67.1 \pm 3.0$ & $0.933 \pm 0.011$ & 0.038 \\
\midrule
A$'$ & qubit brickwall $L=11$, plain           & 432   & $52.1 \pm 0.8$ & $0.850 \pm 0.008$ & 1 \\
D$'$ & qubit $L=11$, full front end       & 1,460 & $54.9 \pm 1.5$ & $0.864 \pm 0.009$ & 0.197 \\
\midrule
Q & free complex map, no physics (baseline) & 1,048,576 & $75.7 \pm 1.8$ & $0.961 \pm 0.004$ & -- \\
\bottomrule
\end{tabular}
\caption{Caltech-10, mean $\pm$ std over three seeds (300 test pictures). Optical rows: four phase plates of
$32\times32$ pixels; qubit rows: brickwall of depth 11 on twelve qubits, D$'$ with the full front end (the circular mask with its phase,
and the mean-centred encoding). Transm.\ (transmission): the
share of the input light that reaches the 128 camera regions (on qubits, the share passing the mask); probabilities
are conditional on it. Q: the encoding, readout and training of row D with a free complex $128\times4096$ map in place
of the mask and plates (Appendix~\ref{app:details}).}
\label{tab:ablation}
\end{table}
\begin{table}[!htb]
\centering\small
\setlength{\tabcolsep}{4pt}
\rowcolors{2}{white}{stripe}
\begin{tabular}{L{0.50\linewidth}rc}
\toprule
model (Caltech-10) & parameters & accuracy (\%) \\
\midrule
qubit brickwall $L=11$ + mask, phase, mean-centred encoding (D$'$) & 1,460 & $54.9 \pm 1.5$ \\
two phase plates of $16\times16$ & 1,540 & 46.7 \\
four phase plates of $16\times16$ & 2,052 & 63.3 \\
two phase plates of $32\times32$ & 3,076 & 59.7 \\
\rowcolor{ours}four phase plates of $32\times32$ (row D) & 5,124 & $72.3 \pm 0.7$ \\
two phase plates of $64\times64$ & 9,220 & 70.7 \\
four phase plates of $64\times64$ & 17,412 & 75.7 \\
\midrule
logistic regression on the raw pixels & 30,730 & 59.7 \\
small convolutional network, same random crops & 25,578 & $85.4 \pm 0.4$ \\
free complex map, no physics (row Q) & 1,048,576 & $75.7 \pm 1.8$ \\
\bottomrule
\end{tabular}
\caption{Size against accuracy on Caltech-10: the full front end with different main circuits, and three references.
Values with $\pm$ are mean and standard deviation over three seeds, the others one seed (row D at that seed: 73.0\,\%).
On colour CIFAR-10 the plate rows reach 33.2, 38.8, 37.2, 48.1, 46.1 and 50.9\,\% in the order listed
(Figure~\ref{fig:params}). ROC-AUC: logistic regression 0.919, convolutional network 0.987.}
\label{tab:sizes}
\end{table}
Table~\ref{tab:ablation} adds one part at a time. The mean-centred encoding is the largest step: $+6.8$ points and ROC-AUC
$0.892 \to 0.939$, well outside the spread of the three seeds, with the same parameters and the same transmission.
A lens mask with amplitude only adds nothing beyond noise and cuts the transmission tenfold; giving the mask a phase
is the second clear step ($+3.4$ points, again outside the seed spread, with the smallest spread between seeds). We
call the mask with its phase plus the mean-centred encoding the \emph{full front end}, and row D the \emph{full
optical model}. Pixel-aware plates keep the accuracy
($71.6 \pm 1.7\,\%$) while making a simulated physical bench give the same answers (LightPipes, ten pictures;
Section~\ref{sec:bench}); a loss term that rewards keeping light doubles the transmission at a cost of 4.4 points
(row F; 4.44 before rounding).

\begin{figure}[!htb]
\centering
\includegraphics[width=\linewidth]{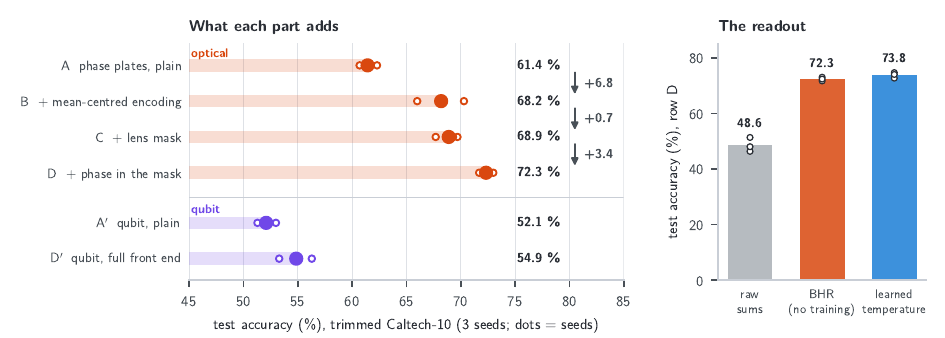}
\caption{Left: what each part adds on Caltech-10 (rows A--D, A$'$ and D$'$ of Table~\ref{tab:ablation}; large dots the mean, small
dots the three seeds). Right: the readout on row D: raw sums, BHR with nothing trained, and a softmax with one
learned temperature.}
\label{fig:ablation}
\end{figure}

The readout matters more than any part before it, and what matters in it is its slope (its temperature). We trained
row D with each readout and its own loss, three seeds each. The plain group sums of the same outcomes (\emph{raw
sums}) give $48.6 \pm 2.5\,\%$ (ROC-AUC $0.859 \pm 0.012$). A softmax of the group sums with one learned temperature
(the parameter of temperature scaling~\cite{guo2017calibration}, here trained with the model) gives $73.8 \pm
1.0\,\%$ ($0.959 \pm 0.004$); its temperature settles at a slope of about 380 per unit of probability. BHR gives $72.3
\pm 0.7\,\%$ ($0.947 \pm 0.008$) with no trained parameter; its slope, $\kappa/(1-b_g) \approx 136$, follows from
$\varphi$. At test time the three readouts rank the classes almost alike (they differ only through the small
differences in group size), so the slope acts during training, where it sets how hard the loss pushes the true class
ahead; slopes of 136 and about 380 both train well, the raw sums' slope of one does not. BHR's gain over the raw sums is
therefore the slope it fixes in closed form, in units of headroom, not an accuracy advantage over a learned
temperature (Figure~\ref{fig:ablation}). The accuracy stays within the seed spread over a fourfold range of slopes, and as trained BHR is underconfident (expected calibration
error 0.10): it fixes the slope for training, not for calibration (the sweep and the calibration of the readouts:
Appendix~\ref{app:checks}).

\begin{figure}[!htb]
\centering
\includegraphics[width=\linewidth]{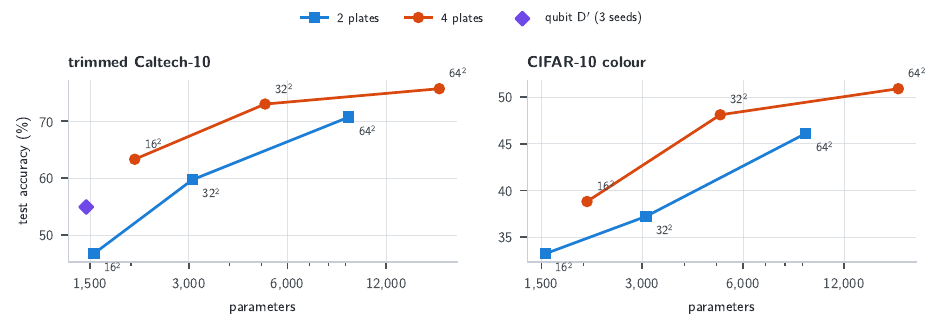}
\caption{Test accuracy against parameters for two and four phase plates of 16, 32 and 64 pixels a side in the
full optical model of row D (Caltech-10; colour CIFAR-10 with the mean-centred encoding; one seed each), with the
qubit model D$'$ for reference.}
\label{fig:params}
\end{figure}

On the same data the plates exceed the qubit brickwall by 9 points with plain amplitudes (A against A$'$) and by 17
points with the full front end (D against D$'$). Part of the second gap is the mask: D$'$ uses the circular mask and D
the lens mask, and a single-seed qubit run with the lens mask (depth 11, trimmed crops) reached 63.7\,\%. The plates also have 3.5--9.5 times the
parameters, but parameters alone do not explain the gap (Table~\ref{tab:sizes}, Figure~\ref{fig:params}): four plates
of $16\times16$ pixels, with a few hundred parameters more than D$'$, are 7 points better at the same seed (63.3 against 56.3\,\%). Depth pays more than pixels:
four plates of $32\times32$ do better than two of $64\times64$ with fewer parameters. Two classical references place
these numbers: the plates sit between a linear classifier of the pixels and a small convolutional network trained with
the same random crops.

\subsection{Four datasets, five settings}
\label{sec:standard}
\begin{table}[t]
\centering\footnotesize
\setlength{\tabcolsep}{2.4pt}
\begin{tabular}{llrcccc}
\toprule
& & & \multicolumn{2}{c}{plain amplitudes} & \multicolumn{2}{c}{mean-centred} \\
\cmidrule(lr){4-5}\cmidrule(lr){6-7}
dataset & pictures & train / test & accuracy (\%) & ROC-AUC & accuracy (\%) & ROC-AUC \\
\midrule
MNIST           & grey $28\times28$   & 58k / 10k   & $96.8 \pm 0.2$ & $0.999 \pm 0.000$ & $96.8 \pm 0.1$ & $0.999 \pm 0.000$ \\
Fashion-MNIST   & grey $28\times28$   & 58k / 10k   & $86.8 \pm 0.3$ & $0.988 \pm 0.000$ & $86.7 \pm 0.3$ & $0.988 \pm 0.000$ \\
CIFAR-10        & grey $32\times32$   & 48k / 10k   & $37.1 \pm 0.7$ & $0.795 \pm 0.003$ & $37.6 \pm 1.2$ & $0.796 \pm 0.007$ \\
CIFAR-10        & colour $32\times32$ & 48k / 10k   & $46.5 \pm 0.1$ & $0.853 \pm 0.001$ & $\mathbf{47.9 \pm 0.2}$ & $\mathbf{0.861 \pm 0.001}$ \\
Caltech-10      & colour $32\times32$ & 2,644 / 300 & $61.4 \pm 0.8^{\dagger}$ & $0.892 \pm 0.005^{\dagger}$ & $\mathbf{72.3 \pm 0.7}$ & $\mathbf{0.947 \pm 0.008}$ \\
\bottomrule
\end{tabular}
\caption{The all-optical classifier on four datasets (CIFAR-10 in grey and in colour), ten classes each: four phase plates with the lens mask and its phase
(row D of Table~\ref{tab:ablation}; 5,121--5,124 parameters, depending on the number of planes), every pixel an amplitude;
mean $\pm$ standard deviation over three seeds (20 epochs on the standard sets, 50 on Caltech-10). $^{\dagger}$Caltech-10's
plain-amplitude run is row A, without the mask.}
\label{tab:standard}
\end{table}
\begin{figure}[!htb]
\centering
\includegraphics[width=\linewidth]{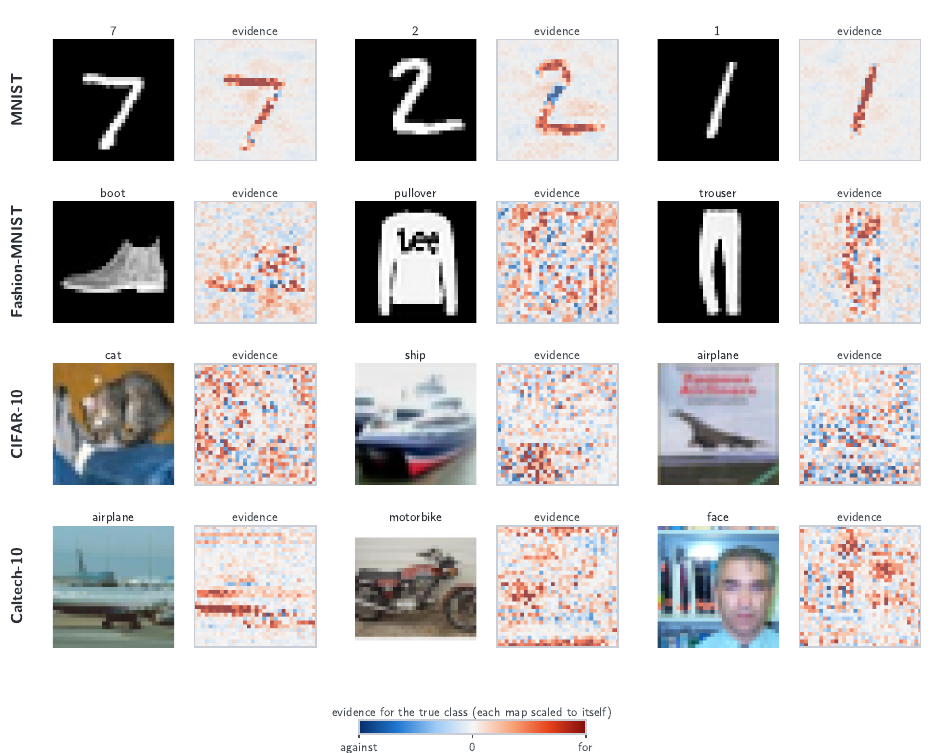}
\caption{Exact evidence for the true class, full optical model (row D, seed 0), on the first three test
pictures of distinct classes in each set (no choice by eye). Red pushes the class above its share, blue pulls it below;
each map is scaled symmetrically to its own 98th percentile of $|e|$, white is zero.}
\label{fig:gallery}
\end{figure}
Table~\ref{tab:standard} gathers the full optical model on all five settings, from handwritten digits to colour
photographs. The mean-centred encoding helps colour CIFAR-10 ($+1.4$ points, outside the spread of the three seeds) and is neutral
on the grey sets (a grey picture's means plane is a constant reference and carries no information about its brightness, Section~\ref{sec:meanplane}). For context only, since architectures, losses and
encodings differ: simulated diffractive networks report 91.75\,\% on MNIST~\cite{lin2018d2nn} and 97.18\,\% with an
improved training~\cite{mengu2019jstqe}, 45.20\,\% on grey CIFAR-10 with $2\times10^5$ neurons (48.51\,\% with
differential detection)~\cite{li2019differential}, 62.13\,\% on grey CIFAR-10 with an ensemble of thirty
networks~\cite{rahman2021ensemble}, and 38.6\,\% on colour CIFAR-10 with three wavelengths~\cite{watanabe2026mo} in a
magneto-optical network~\cite{fujita2022mod2nn}.

Figure~\ref{fig:gallery} shows the evidence of the full optical model on each set. On MNIST the digits stand on a black
background and the evidence follows their strokes; on Fashion-MNIST the pixel maps are noisier, and on CIFAR-10 and
Caltech-10, where the background carries light too, part of the evidence falls on it. The per-pixel maps are speckled,
partly because in coherent propagation neighbouring pixels can contribute with opposite signs.

\subsection{The maps point at what the model uses}
\label{sec:deletion}
\begin{figure}[!htb]
\centering
\includegraphics[width=\linewidth]{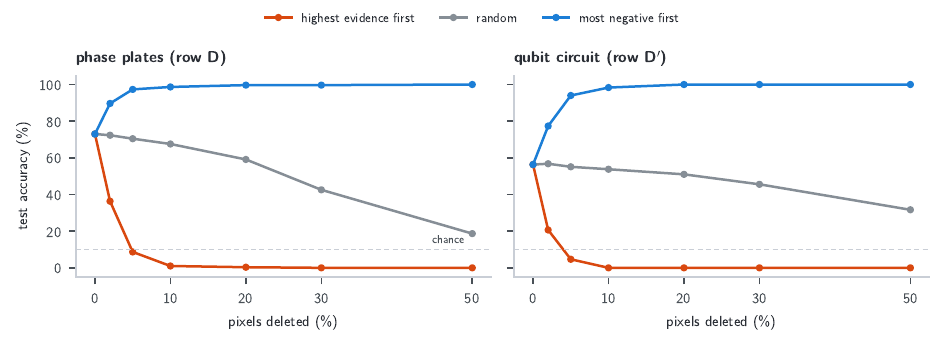}
\caption{Deleting pixels (contrast set to zero) in the order of the exact evidence for the true class, at random, or most
negative first; 300 test pictures, seed 0. Random order: three draws.}
\label{fig:deletion}
\end{figure}

Before the maps are used to judge the model, their faithfulness should be checked. Deleting pixels in the order of the
maps tests this~\cite{petsiuk2018rise}. We set a pixel's contrast to zero in the three colour planes (it becomes the
picture's own mean colour; the means plane is left as it was), renormalise and classify again, ranking the pixels by
the evidence for the true class and, for comparison, by the usual alternatives (Table~\ref{tab:deletion}, Figure~\ref{fig:deletion}). Deleting the 5\,\% of pixels with the
highest evidence drops the plates from 73.0\,\% to 8.7\,\% and the qubit circuit from 56.3\,\% to 4.7\,\%; deleting the
most negative raises them to 97.3\,\% and 94.0\,\%, and the true class's headroom follows the same order. Ranking by the pixel's light alone, which knows nothing of the model, removes far less (55.0\,\% and
47.0\,\% left), so the drop is not that of deleting bright pixels. The plain
gradient and its magnitude remove much less accuracy than the evidence, at small deletions little more than random
order (Figure~\ref{fig:deletion}): a gradient lacks the factor of the pixel's own amplitude. Gradient $\times$ input~\cite{shrikumar2016notjust,
ancona2018towards}, which has that factor, ranks the pixels as well as the evidence (9.0\,\% and 4.3\,\% at 5\,\%),
and must: with the transmission held fixed it is exactly twice the evidence, and the gradient of $P_g$ itself differs
from that only by a term proportional to the attention. Taken through the readout, the gradient $\times$ input of
$\log \hat p_g$, the trained output, removes even more (5.0\,\% and 2.7\,\%). Occlusion, which deletes each pixel
alone and measures the drop, matches the evidence too, because deleting pixel $x$ changes the numerator
$\bra{a}M_g\ket{a}$ by $-2Z\,e_g(x)$ plus a term quadratic in the deleted amplitudes, before renormalisation; it needs 1,024 extra runs per picture, where the
evidence needs one forward and one adjoint pass. This test therefore does not single out the evidence among
attributions that carry the amplitude factor; what the evidence adds is that it sums exactly to $P_g - b_g$ and can be
measured on the hardware (Section~\ref{sec:measure}). The test is also close to the evidence's own definition, since a
deleted pixel is set to the zero baseline of the Integrated Gradients, and deleted pictures leave the training
data~\cite{hooker2019benchmark}. Retraining on deleted pictures (ROAR, the remedy proposed there) cannot rank a class-specific map here: the pattern
of the deleted pixels alone gives the class away (Appendix~\ref{app:checks}). It uses the true label to choose the pixels, so the rise under reversed order shows what the evidence for the
true class holds, not a way for the model to correct itself. That the maps add up exactly and that deleting by them
changes the answer are separate properties; the first is proved, the second is tested here. Ranking instead by the
evidence for the model's own answer, which needs no label, tells the same (Table~\ref{tab:delpred}): a few percent of
the pixels change most answers, wrong answers change sooner than right ones, and deleting the most negative evidence
for the answer never changes it and widens its lead.
\begin{table}[!htb]
\centering\small
\setlength{\tabcolsep}{4pt}
\begin{tabular}{lcccccccc}
\toprule
& \multicolumn{4}{c}{plates (row D), 73.0\,\%} & \multicolumn{4}{c}{qubits (D$'$), 56.3\,\%} \\
\cmidrule(lr){2-5}\cmidrule(lr){6-9}
ranked by \hfill deleted: & 2\,\% & 5\,\% & 10\,\% & 20\,\% & 2\,\% & 5\,\% & 10\,\% & 20\,\% \\
\midrule
\rowcolor{ours}evidence (ours)   & 36.3 & 8.7 & 1.0 & 0.3 & 20.7 & 4.7 & 0.0 & 0.0 \\
occlusion & 36.0 & 9.3 & 1.7 & 0.3 & 21.0 & 4.3 & 0.3 & 0.0 \\
gradient $\times$ input & 37.0 & 9.0 & 1.3 & 0.3 & 20.3 & 4.3 & 0.3 & 0.0 \\
gradient $\times$ input, $\log\hat p_g$ & \textbf{31.3} & \textbf{5.0} & \textbf{0.7} & 0.3 & \textbf{16.0} & \textbf{2.7} & 0.0 & 0.0 \\
gradient                         & 68.7 & 65.0 & 60.0 & 53.0 & 55.0 & 52.3 & 48.3 & 45.0 \\
$|$gradient$|$ (saliency) & 67.3 & 62.0 & 57.7 & 41.7 & 51.7 & 45.7 & 44.3 & 39.7 \\
input intensity & 65.7 & 55.0 & 36.7 & 18.0 & 51.3 & 47.0 & 40.0 & 35.0 \\
random & 72.2 & 70.6 & 68.2 & 59.2 & 56.0 & 55.6 & 54.1 & 50.7 \\
\midrule
evidence, reversed & 89.7 & 97.3 & 98.7 & 99.7 & 77.3 & 94.0 & 98.3 & 100.0 \\
\bottomrule
\end{tabular}
\caption{Deletion test: test accuracy (\%) after deleting the given share of pixels, highest first in each ranking
(contrast set to zero, renormalised; 300 test pictures, seed 0). Occlusion deletes each pixel alone first (1,024
extra runs per picture); gradient $\times$ input: the pixel's amplitudes times the gradient of $P_g$, or of the
readout's $\log\hat p_g$, summed over the colour planes; input intensity: the pixel's encoded light; random: three draws; \emph{reversed}: the most negative
evidence first; bold: the lowest per column where the rows differ. Lower is better for
every row but the last.}
\label{tab:deletion}
\end{table}
\begin{table}[!htb]
\centering\small
\setlength{\tabcolsep}{4pt}
\begin{tabular}{lccccccc}
\toprule
& \multicolumn{4}{c}{answers kept (\%) after deleting} & \multicolumn{2}{c}{kept after 2\,\%} & lead \\
\cmidrule(lr){2-5}\cmidrule(lr){6-7}
& 2\,\% & 5\,\% & 10\,\% & 20\,\% & right & wrong & before $\to$ after 5\,\% \\
\midrule
plates (row D) & 39.3 & 9.3 & 1.0 & 0.3 & 49.8 & 11.1 & $+0.016 \to -0.018$ \\
qubits (D$'$) & 26.0 & 4.7 & 0.0 & 0.0 & 36.7 & 12.2 & $+0.009 \to -0.016$ \\
\bottomrule
\end{tabular}
\caption{Deletion ranked by the evidence for the model's own answer (no label used; same deletion as
Table~\ref{tab:deletion}, seed 0, 300 test pictures). \emph{Right, wrong}: among the pictures first answered right
(219 plates, 169 qubits) or wrong. \emph{Lead}: the answer's headroom minus the largest other headroom, mean over the
pictures. Deleting the most negative evidence first keeps every answer at every fraction.}
\label{tab:delpred}
\end{table}

\subsection{The explanations sit on the object}
\label{sec:object}
With the deletion test supporting their ranking, the maps can judge the encoding. Caltech-101 provides an outline of the object in every
picture, which we carry through the trim and the crop onto the $32\times32$ grid. The outline covers 41\,\% of a crop on
average. Table~\ref{tab:object} compares where the maps put their weight. With plain amplitudes the evidence is spread
evenly (1.00--1.04 times the outline's area share, its peak inside the outline in under half of the pictures). With the
mean-centred encoding it concentrates on the object: 1.32--1.37 times the area share for the plates (peak inside in 68--71\,\% of
the pictures) and 1.26 for the qubit circuit (62\,\%). The shift comes from the encoding: the input intensity already sits 1.42 times on the
object, and the trained circuits keep it there. With every trained value of the plates and the mask permuted at random
(a variant of the model-parameter randomisation test~\cite{adebayo2018sanity}, permuting the values rather than
re-initialising them), the positive evidence still sits 1.18 times on the object
(peak inside in 61\,\% of the pictures; the qubit model 1.18 and 59\,\%), so the trained plates add the rest, to 1.37,
and the trained circuit little. The signed maps do change with the model: those of the permuted models correlate with
the trained ones at 0.00 when the whole model is permuted (0.02--0.06 when only its top part is; median Spearman over
pictures), their magnitudes at 0.33--0.35, and the magnitude
follows the pixel's light about equally in trained and permuted models (0.54--0.60). Summaries of where the magnitude
or the positive part lies, such as Table~\ref{tab:object}, therefore mostly describe the encoding; the signs carry
the model. The ketch in
Figure~\ref{fig:overview} (and in Figure~\ref{fig:hwmaps}) shows the encoding's part: the evidence and the attention are
strongest on the dark hull at the waterline, the part that plain amplitudes would barely light.
\begin{table}[t]
\centering\small
\begin{tabular}{lccc}
\toprule
row (Table~\ref{tab:ablation}) & input intensity / area & positive evidence / area & its peak inside \\
\midrule
A  plates, plain           & 1.02 & 1.04 & 0.49 \\
B  + mean-centred           & 1.42 & 1.32 & 0.68 \\
\rowcolor{ours}D  full optical model      & 1.42 & 1.37 & 0.71 \\
A$'$ qubit, plain          & 1.02 & 1.00 & 0.42 \\
D$'$ qubit, full front end & 1.42 & 1.26 & 0.62 \\
\bottomrule
\end{tabular}
\caption{Share of each map inside the Caltech-101 object outline divided by the outline's area share (1 = spread
evenly), and the share of pictures whose strongest pixel lies inside it. Means over three seeds, 300 test pictures.}
\label{tab:object}
\end{table}

\subsection{The maps find and fix a shortcut}
\label{sec:shortcut}
\begin{figure}[!htb]
\centering
\includegraphics[width=\linewidth]{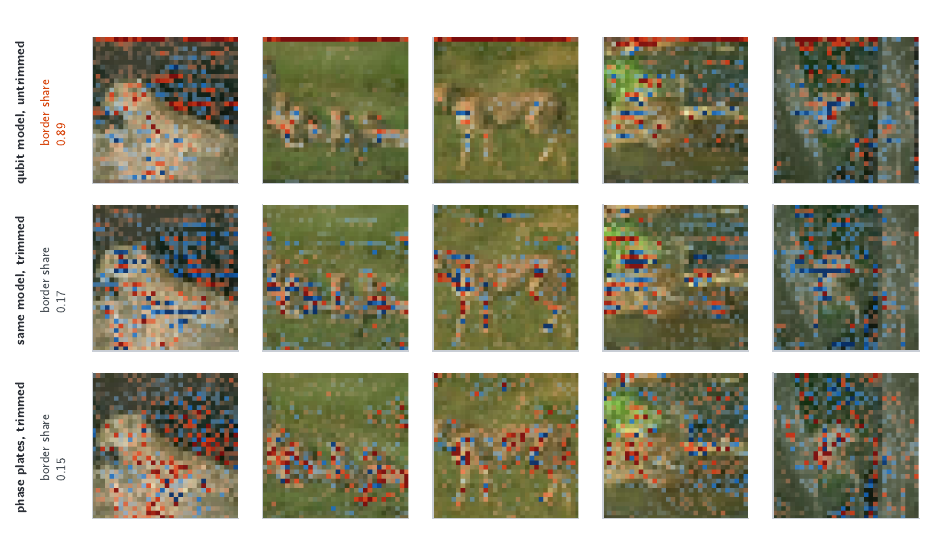}
\caption{Evidence for leopard over five test pictures (red for the class, blue against it; faint where near zero): the qubit model (depth 11, lens mask) on untrimmed crops (top), the
same model on trimmed crops (middle) and the phase plates (bottom). Border share: median share of the net evidence
in the 2-pixel border ring (23\,\% of the area).}
\label{fig:leopard}
\end{figure}

The maps can also find a flaw in the data. On the untrimmed crops the maps of a qubit model (depth 11 with the lens
mask) showed the leopard's evidence running along the top edge of the picture. The cause is in the data: leopard
pictures carry a dark top row (mean brightness 50 against 88 one row lower, 150 for the other classes). For leopards,
a median 0.89 of the net evidence lay in the 2-pixel border ring, which covers 23\,\% of the area. Blanking the ring to
each picture's own mean dropped that model's leopard accuracy to 10\,\% and its overall accuracy from 62\,\% to
47\,\%. Trimming 4\,\% of every side removed the frame. Retrained on the trimmed set (now also with random crops), the
same architecture puts a median 0.17 of the leopards' evidence on the ring, and the phase plates put 0.15, both below
the ring's area share; the evidence moves onto the body and legs (Figure~\ref{fig:leopard}). The maps also explain
errors: a yellow biplane with a round marking (not shown) is called a leopard, and its leopard evidence lies on the
marking and the dark undercarriage, a spotted yellow-and-black texture.

How the shortcut was found bears on the test set, so we state it. The edge was seen in the maps of test pictures, the
statistics above were measured on the test set, and the trim was then set and applied to the same split, which was
kept. The test set therefore informed one preprocessing decision. Model selection used only the 100 validation
pictures, and $\varphi$ was fixed before these experiments, but no untouched holdout of Caltech-10 remains (the trimmed set uses every picture of the ten classes): the
results on the trimmed set carry this caveat. A retrospective analysis of the training pictures independently
supports the presence of the border shortcut. On the 2,644 training pictures the leopards' top row has a mean brightness of 54 against 92 one row lower (168 for the
other classes), and the untrimmed model's maps put a median 0.98 of the leopards' net evidence in the border ring,
the highest of any class (hawksbill is next, at 0.45); the 100 validation pictures, four of them leopards, agree.
This supports the rationale for trimming, but does not undo the test set's role in the original preprocessing
decision.

\subsection{On a quantum computer}
\label{sec:hardware}
\begin{figure}[!htb]
\centering
\includegraphics[width=\linewidth]{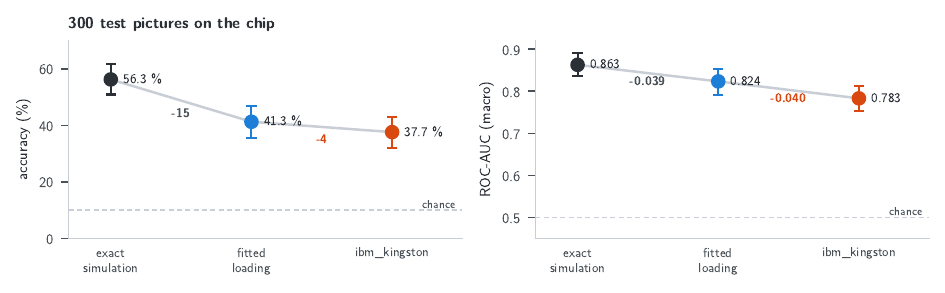}
\caption{The whole test set on hardware, step by step: exact simulation, the fitted loading circuit without noise, and
ibm\_kingston (300 pictures, 95\,\% bootstrap intervals). Confusion matrices and ROC curves: Appendix~\ref{app:hardware}.}
\label{fig:ibmsteps}
\end{figure}

We ran all 300 Caltech-10 test pictures through model D$'$ (seed 0) on IBM's ibm\_kingston (Heron r2), one circuit per
picture, 4,000 shots, in ten jobs of thirty pictures (340 quantum seconds). For each picture, the masked state is
computed off the chip (on a bench, the 4f system in front of the plates), so the chip runs a hybrid model whose trained
linear front end, the mask, is classical; the state is loaded by a fitted RY--CNOT circuit of
depth 8, with 44 CNOTs (105 with the trained circuit) and a median state fidelity of 0.77; an exact preparation would need about 3,900 CNOTs. The
trained depth-11 brickwall follows without any SWAP. Accuracy falls in steps we can measure separately
(Figure~\ref{fig:ibmsteps}): from 56.3\,\% exact to 41.3\,\% with the fitted loading and to 37.7\,\% on the chip, 3.8 times
chance; ROC-AUC $0.863 \to 0.824 \to 0.783$ (confusion matrices and ROC curves in Figure~\ref{fig:hwcm}). Loading the picture costs 15 points; the chip's noise costs another 4. Paired over the same pictures (exact
McNemar tests), the loss to loading is significant ($p < 10^{-9}$), the further loss on the chip is not ($p = 0.19$; 35
pictures lost, 24 gained). Model A$'$, the qubit circuit on plain amplitudes with no mask and so no classical front end, ran the same way on
the same pictures and ends close to D$'$ on the chip (32.3\,\%, paired $p = 0.16$), but loses in a different step: its
plain states load almost without loss, and the chip costs it 22 points (Appendix~\ref{app:checks}).
Every job of the test-set and map runs, with its calibration, the qubits used, the
circuits as sent and the raw counts, is kept in a ledger released with the code.

\emph{Maps of pictures drawn at random.} After the test-set job and before any map circuit was fitted or run we
fixed 30 test pictures, three per class drawn with a recorded seed among that class's 30, with no regard to the chip's
answers (it had classified 13 of them correctly), and four of these for $4\times4$ maps; the selection and both
estimators, with neither named as the main one, are released with the code. We lead with the sign flip; the
noisy simulation of the same circuits orders the two the same way (0.90 against 0.80 at 4,000 shots;
Table~\ref{tab:hwmaps}). For each of a picture's
four blocks its job measured $\psi_S$, $\psi_{\bar S}$ and $\psi_S - \psi_{\bar S}$, and the whole picture in the same
job, 13 circuits a picture, so both estimators (Eqs.~\eqref{eq:polar} and \eqref{eq:flip}) come from the same counts: six
jobs of five pictures, 4,000 shots, 426 quantum seconds. Table~\ref{tab:hwmaps} compares the chip's maps with the
exact ones over the 120 blocks. The exact maps are those of the ideal masked state, so the chip's maps also carry the
error of the fitted block preparations (state fidelity median 0.78). They are maps of the filtered state, which
stands for the picture only as far as the mask's kernel reaches (Appendix~\ref{app:checks}).

Four baselines show what the agreement is worth. Spreading each picture's exact $P_g - b_g$ evenly over its four
blocks, a map with no spatial information, correlates at 0.31, because the pooled correlation also rewards getting each
picture's total right; spreading it in proportion to each block's light $\|\psi_S\|^2$, the null closest to the
estimator's own amplitude weighting, correlates at 0.42 (centred 0.37); permuting each chip map's blocks within its
picture gives correlations below 0.29 in 95\,\% of draws; and 73\,\% of the exact blocks are positive, so calling every
block positive matches 73\,\% of the signs, against the chip's 88\,\%. With each picture's mean block removed
from both maps, which leaves the spatial pattern alone, the correlation is 0.93. The chip's maps agree as well for the
17 pictures it classified wrongly ($r = 0.87$) as for the 13 it classified correctly (0.89).

A correlation shows neither the size of the evidence nor the sum rule, so we also compare the values. With
Eq.~\eqref{eq:flip} the chip's blocks have a mean absolute error of 0.0025, where the mean $|E_S|$ is 0.0062 (relative
error 0.41), 1.5 times the error of shot noise alone (0.0017, drawn at 4,000 shots from the exact distributions). A fit
through zero has slope 0.82; the noisy simulation shows the same shrinkage (slope 0.90), as gate noise pulls each class
probability towards $b_g$. The four blocks of a map add up to the same job's measured $P_g - b_g$ within 0.0035 on
average (mean absolute residual; with no offset, the signed mean being $-0.0003$), and to the exact $P_g - b_g$ (mean 0.016) within 0.0071, where shot noise alone gives 0.0050.
Gate and readout noise that do not depend on the prepared state leave each class term a quadratic form of the input
(a depolarising channel, for instance, turns $M_g$ into $(1-p)M_g$ plus a multiple of the identity), so the identity
and the sum rule against the job's own whole-picture term still hold; only errors that depend on the state break them,
mainly the fitted loading. Against the noisy
device's own maps, those of the noisy simulation of the same circuits without shot noise, the chip's maps correlate at
0.93, so what remains is mostly shot noise (the share of each step: Appendix~\ref{app:checks}). No readout-error mitigation was used; being
linear, it would keep the identity.
\begin{figure}[!htb]
\centering
\includegraphics[width=\linewidth]{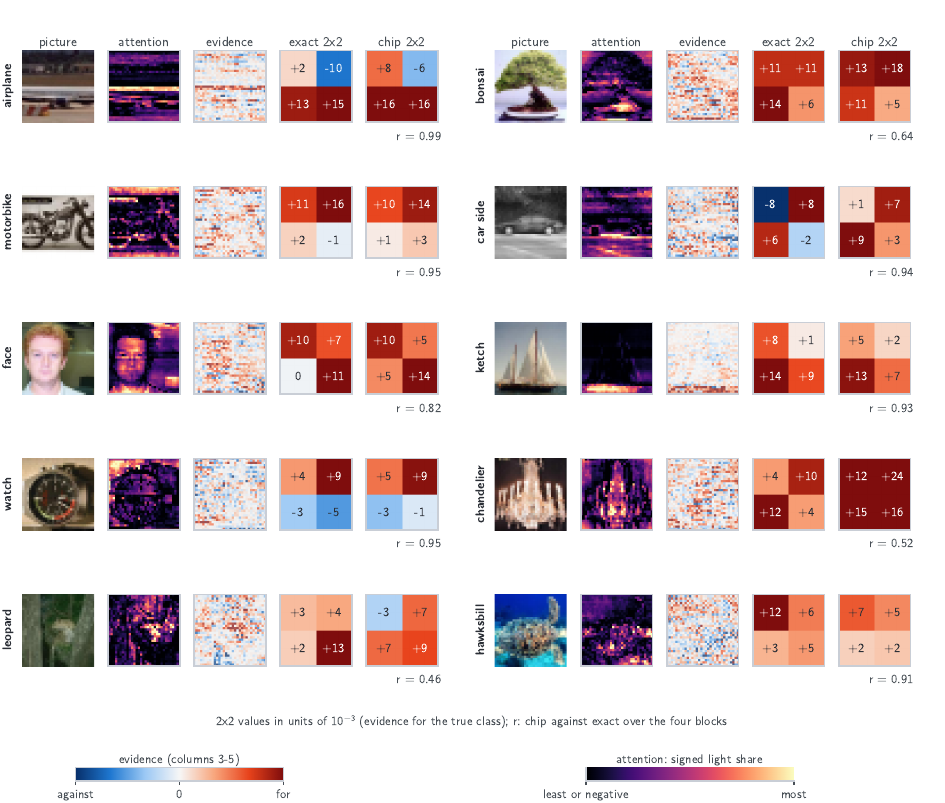}
\caption{The earlier maps: ten test pictures, one per class, chosen by the rule of Section~\ref{sec:hardware}. Columns: the picture; the exact attention, each pixel's signed share of the light
that reaches the measurement against the average pixel's (dark: little or negative, bright: most); the exact evidence per
pixel; the exact $2\times2$ evidence; the $2\times2$ evidence measured on ibm\_kingston. The pixel maps are over the
picture; the $2\times2$ maps over blocks of the filtered state the chip receives. Evidence: red for the true
class, blue against, white zero, each pixel map scaled to its own 98th percentile of $|e|$ and each pair of $2\times2$
maps to the larger exact value.}
\label{fig:hwmaps}
\end{figure}

\begin{table}[!htb]
\centering\small
\setlength{\tabcolsep}{4pt}
\begin{tabular}{lcccc}
\toprule
$2\times2$ maps against the exact ones & correlation & centred & same sign & strongest block \\
\midrule
\multicolumn{5}{l}{\emph{30 pictures drawn at random (120 blocks)}}\\
ibm\_kingston, Eq.~\eqref{eq:flip} & 0.88 (0.84--0.91) & 0.93 & 88\,\% & 23 / 30 \\
ibm\_kingston, Eq.~\eqref{eq:polar} & 0.82 (0.76--0.87) & 0.92 & 84\,\% & 21 / 30 \\
noisy simulation, 4,000 shots, Eq.~\eqref{eq:flip} & 0.90 & 0.95 & 82\,\% & \\
noisy simulation, 4,000 shots, Eq.~\eqref{eq:polar} & 0.80 & 0.94 & 79\,\% & \\
\midrule
\multicolumn{5}{l}{\emph{10 pictures chosen by a rule (40 blocks), earlier maps}}\\
ibm\_kingston, Eq.~\eqref{eq:polar} & 0.77 (0.61--0.91) & 0.85 & 88\,\% & 7 / 10 \\
\bottomrule
\end{tabular}
\caption{Evidence maps measured on the chip against the exact ones. Brackets: 95\,\% bootstrap intervals over pictures.
Centred: each picture's mean block removed from both maps. Strongest block: pictures whose largest measured block is
the largest exact one. The noisy simulation runs the same circuits with the backend's noise model and draws the shots.}
\label{tab:hwmaps}
\end{table}

\emph{$4\times4$ maps} of three of the four pictures fixed for them (33 circuits each) correlate with the exact maps
at 0.60, but at 4,000 shots a $4\times4$ block is at the shot-noise limit (Appendix~\ref{app:checks}).

\emph{The earlier maps}, measured before, used ten pictures chosen by a stated rule: one per class, classified correctly
on the chip with a BHR probability $\hat p_g$ of at least 0.6 for the true class in the test-set job, and of those the
one whose background covers most of the crop and varies most (motorbike had no such picture and took its most
confident correct one; three classes had a single candidate). Each $2\times2$ map took nine circuits
(Eq.~\eqref{eq:polar}), all 90 in one job, with each picture's whole-picture circuit; they appear in
Figure~\ref{fig:hwmaps}. Their agreement (Table~\ref{tab:hwmaps}) is no better than the random pictures', so the choice
did not flatter them. A first run, which reused each picture's whole-picture term from the test-set job, gave 0.63 (Appendix~\ref{app:checks}). In simulation, with exact
states and 4,000 shots, the measurement recovers the exact maps of all 300 test pictures with a pooled correlation of
0.91, and 0.88--0.89 for the pictures the chip classified wrongly, for those it classified correctly with probability
below 0.6, and for the half with the smaller exact $P_g - b_g$.
\emph{A deliberate fault.} In simulation, every fault we put into the trained circuit that moves the maps beyond the
chip's error at 4,000 shots also costs the noise-free model at least 9.7 points of accuracy (Appendix~\ref{app:checks}),
so we did not spend chip time on asking whether a measured map shows a fault that the accuracy does not.

\subsection{On a simulated lens bench}
\label{sec:bench}
We rebuilt the trained pixel-aware model of row E in LightPipes, an independent scalar-optics package (details in Appendix~\ref{app:bench}): each plane
through its own 4f system with the learned mask, the planes tiled, Fresnel propagation through the plates with square
pixels, the light summed over the camera regions. The output of each 4f system reaches the first plate as it is,
smooth and band-limited, sampled four times per pixel. Table~\ref{tab:lightpipes} compares the light per detector
region with our model on ten test pictures, by its overlap (the normalised inner product of the two vectors of region
intensities) and by the prediction. Two of its rows test what both simulators assume: glass plates of real thickness
in place of thin phase masks, and LightPipes' convolution propagator in place of its spectral one. The thin-plate
approximation holds at these 16\,\textmu m pixels. Under the convolution propagator the answers change for two pictures, both with $\hat p_g$ below 0.36;
this grid samples the field at $\Delta x = \lambda d/L$, where the two methods are about equally accurate. Both are
paraxial; at the system's numerical aperture of about 0.04 the paraxial phase error, $kd\,\mathrm{NA}^4/8 \approx
0.01$\,rad, is a tenth of the plate phase noise the model tolerates (Table~\ref{tab:robust}).
\begin{table}[!htb]
\centering\small
\setlength{\tabcolsep}{4pt}
\rowcolors{2}{white}{stripe}
\begin{tabular}{L{0.62\linewidth}cc}
\toprule
LightPipes bench (10 test pictures) & overlap (min) & same prediction \\
\midrule
row E, 4f output passed on as it is & 0.997 (0.996) & 10 / 10 \\
row E, 4f output re-made into square pixels & 0.998 (0.996) & 10 / 10 \\
row E, glass plates ($n = 1.457$) crossed in 32 layers & 0.997 (0.9945) & 10 / 10 \\
row E, convolution propagator instead of the spectral one & 0.995 (0.993) & 8 / 10 \\
row D, point-sampled plates, 4f output as it is & 0.980 (0.971) & 7 / 10 \\
\bottomrule
\end{tabular}
\caption{The trained models rebuilt in LightPipes, against our model: overlap of the light per detector region (mean
over the pictures, minimum in brackets) and the share of identical predictions (95\,\% interval for 10 / 10:
0.69--1). Row D's plates were trained point-sampled; the difference is the diffraction of square pixels, which the
pixel-aware training of row E includes.}
\label{tab:lightpipes}
\end{table}
\begin{figure}[!htb]
\centering
\includegraphics[width=\linewidth]{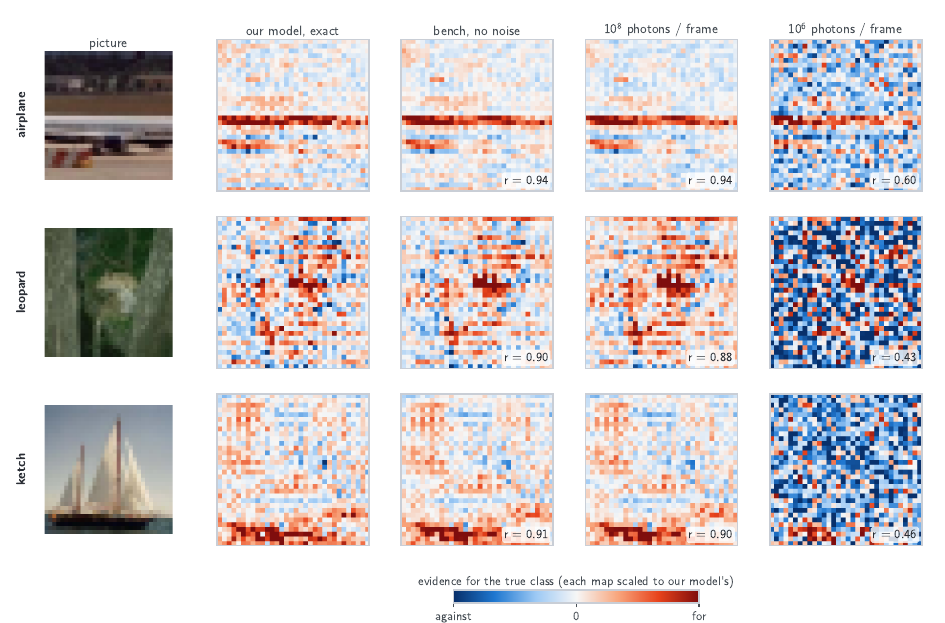}
\caption{Per-pixel evidence measured on a simulated lens bench (LightPipes) from camera intensities only
(Eq.~\eqref{eq:bench}), for the pixel-aware model of row E: our model's exact map, the bench without noise, and with
Poisson noise of $10^8$ and $10^6$ photons per camera frame. Our model's map counts the means plane at each pixel, as
the bench's blocks do. All maps of a row share one scale, our model's 98th
percentile of $|e|$; white is zero.}
\label{fig:bench}
\end{figure}
\begin{table}[!htb]
\centering\small
\setlength{\tabcolsep}{4pt}
\rowcolors{2}{white}{stripe}
\begin{tabular}{L{0.36\linewidth}cccc}
\toprule
per-pixel map ($32\times32$) & airplane & leopard & ketch & ketch's sum \\
\midrule
\multicolumn{5}{l}{\emph{correlation with our model's exact map}} \\
bench, no noise & 0.94 & 0.90 & 0.91 & 0.038 \\
$10^8$ photons per frame & 0.94 & 0.88 & 0.90 & \\
$10^6$ photons per frame & 0.60 & 0.43 & 0.46 & \\
\midrule
\multicolumn{5}{l}{\emph{correlation with the flawless bench's map}} \\
laser drift 0.1\,\% per frame & 0.996 & 0.998 & 0.984 & 0.042 \\
laser drift 1\,\% per frame & 0.71 & 0.80 & 0.49 & $-0.141$ \\
off pixels pass 1\,\% of their amplitude & 1.000 & 1.000 & 1.000 & 0.040 \\
off pixels pass 3\,\% & 1.000 & 1.000 & 1.000 & 0.056 \\
crosstalk 2\,\% to each off neighbour & 0.999 & 0.999 & 0.999 & 0.038 \\
crosstalk 5\,\% & 0.996 & 0.993 & 0.994 & 0.038 \\
\bottomrule
\end{tabular}
\caption{Evidence for the true class measured on the LightPipes bench from camera intensities only (row E, three test
pictures). Our model's map is compared like for like: at each pixel it includes the means plane's amplitude at that
place, as the bench's block (the same pixels in all four tiles) does. \emph{Ketch's sum}: the measured map's sum, which should equal the bench's own $P - b = 0.038$; drift and
leakage move it, crosstalk does not. Drift: a random relative error of the laser power per frame. Crosstalk: an edge
pixel lends this fraction of its amplitude to each \emph{off} neighbour.}
\label{tab:benchmaps}
\end{table}

We then measured evidence maps on this simulated bench by Eq.~\eqref{eq:bench}, from camera intensities alone, at the
full resolution of $32\times32$ pixels (2,049 frames per picture; Figure~\ref{fig:bench}, Table~\ref{tab:benchmaps}).
For each of three pictures the measured map adds up to the bench's own $P - b$ to $5\times10^{-15}$, so the procedure returns
the simulated bench's own decomposition. It resembles our model's map (correlation 0.90--0.94, with the model's map
also counting the means plane at each pixel, as the bench's blocks do); the residual is the model's own
difference from the bench, which also moves $P - b$ (for the airplane 0.026 on the bench against 0.033 in the model). On
ten further test pictures, one per class drawn at random (seed fixed beforehand), the maps add up to the bench's own
$P - b$ to $10^{-14}$ and correlate with our model's at 0.88--0.93 (median 0.90; 0.87--0.92 with $10^8$ photons per
frame), so the three pictures of Table~\ref{tab:benchmaps} are typical. With
$10^8$ photons per frame the maps correlate with our model as well as the noise-free ones do; $10^6$ are too few. A map
needs more light than a classification ($10^5$ photons per picture suffice to classify, Section~\ref{sec:builders}),
because it is built from differences of frames. A milliwatt laser gives far more than $10^8$ photons per 10-ms frame;
the practical limit is the camera's full-well capacity and dynamic range, which may call for averaging frames.

Three flaws of a real bench were then applied to the measurement (Table~\ref{tab:benchmaps}, lower part). Laser power
that drifts between frames matters most: the laser must hold its power to about 0.1\,\% over a map, or each frame be
normalised by a reference detector. A modulator whose \emph{off} pixels still pass some amplitude leaves the map's
pattern unchanged but scales it, so the map no longer adds up to $P - b$. Crosstalk between modulator pixels changes the
maps little and keeps their sum. The sum rule $\sum_S E_S = P - b$ is thus a ready self-test on a bench: it flags
drift and leakage, and crosstalk does little harm.

\subsection{For builders}
\label{sec:builders}
\begin{figure}[!htb]
\centering
\includegraphics[width=\linewidth]{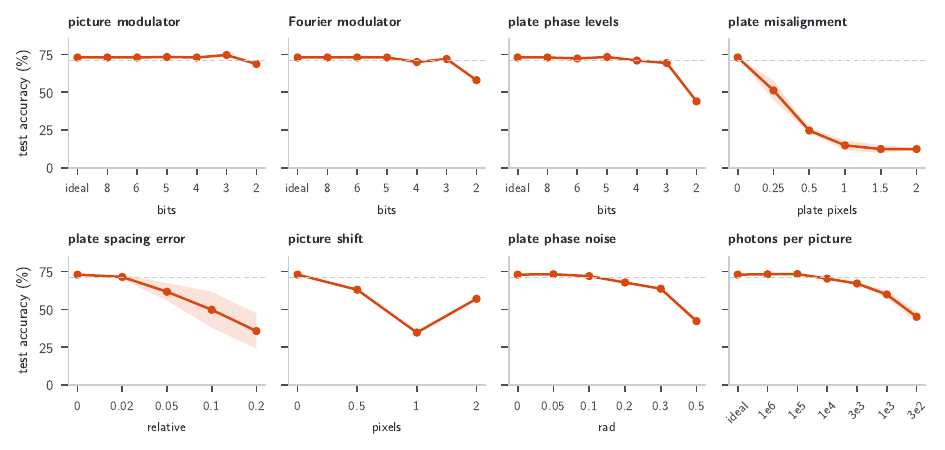}
\caption{Test accuracy of the pixel-aware model (row E, seed 0) under bench flaws, applied without retraining. Three draws each; shaded: spread over the draws; dashed: 2 points below the aligned
model (the tolerance of Table~\ref{tab:robust}).}
\label{fig:robust}
\end{figure}
\begin{table}[!htb]
\centering\small
\setlength{\tabcolsep}{4pt}
\rowcolors{2}{white}{stripe}
\begin{tabular}{L{0.27\linewidth}L{0.21\linewidth}L{0.46\linewidth}}
\toprule
flaw & tolerated & breaks \\
\midrule
picture modulator levels & 3--4 bits & 2 bits: 68.7\,\% \\
mask modulator levels & 5 bits & 4 bits: 70.0\,\%; 3 bits: 72.0\,\%; 2 bits: 58.0\,\% \\
plate phase levels & 5--6 bits & 4 bits: 71.0\,\%; 2 bits: 44.0\,\% \\
plate sideways shift & 0.05 plate px (0.8\,\textmu m; 72.9\,\%) & 0.1 px: 69.2\,\%; 0.25 px: 51.2\,\%; 0.5 px: 24.7\,\% \\
plate rotation & 0.25$^\circ$ (71.1\,\%) & 0.5$^\circ$: 66.0\,\%; 1$^\circ$: 58.9\,\%; 2$^\circ$: 41.9\,\% \\
mask off-centre & 0.25 mask px (2\,\textmu m; 71.7\,\%) & 0.5 px: 63.6\,\%; 1 px: 38.2\,\% \\
4f magnification error & 0.5\,\% (72.7\,\%) & 1\,\%: 69.0\,\%; 2\,\%: 60.3\,\% \\
camera regions shifted & 1 px (8\,\textmu m; 71.8\,\%) & 2 px: 66.8\,\% \\
plate spacing error & 2\,\% (71.4\,\%) & 5\,\%: 61.7\,\%; 10\,\%: 49.8\,\% \\
picture shift, whole field & $<$ 0.5 px & 0.5 px: 63.0\,\%; 1 px: 34.7\,\%; 2 px: 57.0\,\% \\
picture shift, means fixed & 0.5 px (73.0\,\%) & 1 px: 68.7\,\%; 2 px: 62.0\,\%; 4 px: 35.3\,\% \\
plate phase noise & 0.1 rad (72.0\,\%) & 0.2 rad: 67.8\,\%; 0.5 rad: 42.2\,\% \\
photons per picture & $10^5$ (73.3\,\%) & $10^4$: 70.3\,\%; $10^3$: 59.9\,\% \\
\bottomrule
\end{tabular}
\caption{Bench flaws applied to the trained pixel-aware four-plate model on Caltech-10 (row E, seed 0, 73.0\,\%
aligned), three random draws per flaw. \emph{Tolerated}: the coarsest setting tested that, with every finer one, loses less than 2 points. \emph{Mask modulator}: the
Fourier-plane modulator; \emph{spacing error}: per gap; \emph{phase noise}: random, on the plates; \emph{photons}: at the
camera; \emph{whole field}: the means tile shifted too; \emph{means fixed}: the means tile held in place;
\emph{rotation}: each plate about its centre; \emph{magnification}: the 4f image scaled against the plates. Shifts and
rotations in random directions.}
\label{tab:robust}
\end{table}
We applied bench flaws to the trained pixel-aware model without retraining it (Figure~\ref{fig:robust},
Table~\ref{tab:robust}). Modulators of 8 bits are enough everywhere: the picture tolerates 3--4 bits, the Fourier-plane
modulator 5 and the plates 5--6. A 2\,\% error in plate spacing or a random phase error of 0.1\,rad costs one to two
points, and $10^5$ photons per picture at the camera suffice. The hard requirement is alignment: each plate must sit
within about a twentieth of a plate pixel sideways (0.8\,\textmu m; a tenth costs four points, a quarter drops
accuracy to 51\,\%) and turn by no more than about a quarter of a degree; the Fourier-plane mask must be centred to
about 2\,\textmu m and the 4f magnification right to 0.5\,\%, while the camera regions may be off by a whole
pixel (Table~\ref{tab:robust}). Training with random shifts, as proposed for misalignment-resilient diffractive networks~\cite{mengu2020misalign},
buys tolerance at a cost of
2--3.4 points of aligned accuracy (one seed; separate models trained with the light-keeping term, Table~\ref{tab:shifttrain}).
A model trained with shifts of up to half a plate pixel keeps 47\,\% instead of 23\,\% when a plate is 8\,\textmu m off.
Four plates 3.2\,mm apart are closer than separate modulators can be mounted: they can be fabricated as fixed phase
plates, or one reflective modulator can carry all four with a mirror folding the light between them, as in
multi-plane light conversion~\cite{labroille2014mplc}; the aperture after each plate becomes an absorbing stop, and
anti-reflection coatings keep the coherent light from forming fringes between the plates. A first test of the
camera-only measurement needs less: a modulator for the picture, a single trained layer and a camera, as in a
reconfigurable diffractive processor, whose trained layer is a second, phase-only modulator~\cite{zhou2021dpu}. Much of the means plane's
light lies at the edge of the band, $|k| = 16$, because its tile repeats every two pixels; a mask off-centre in the
Fourier plane cuts it first.
Shifting the whole field, means tile included, dips at one pixel and partly recovers at two, because the means tile
repeats every two pixels and an odd shift swaps its colour cells; with the means tile held in place, as a separate
reference on a bench would be, the dip is gone (Table~\ref{tab:robust}).
\begin{table}[!htb]
\centering\small
\begin{tabular}{lccc}
\toprule
trained with shifts up to (plate, picture) & aligned & plate off 0.25\,px & plate off 0.5\,px \\
\midrule
none                       & 68.7 & 54.0 & 23.1 \\
0.25 plate px, 0.5 px      & 66.7 & 58.3 & 32.7 \\
0.5 plate px, 1 px         & 65.3 & 60.9 & 46.6 \\
\bottomrule
\end{tabular}
\caption{Training with random shifts: test accuracy (\%) of separate models (seed 0, with the light-keeping term, row F
settings), aligned and with each plate shifted sideways.}
\label{tab:shifttrain}
\end{table}

\section{Discussion: how this compares}
\label{sec:discussion}
Table~\ref{tab:compare} sets this work beside the works it builds on, on the qubit side and on the optical side. The
comparison is of kind, not a ranking: datasets, encodings, losses and budgets differ, and most optical numbers are
simulations, as ours are.

\begin{table}[!htb]
\centering\footnotesize
\setlength{\tabcolsep}{3pt}
\renewcommand{\arraystretch}{1.15}
\begin{tabular}{L{0.21\linewidth}L{0.14\linewidth}L{0.19\linewidth}L{0.20\linewidth}L{0.19\linewidth}}
\toprule
work & classical network? & data & reported result & explanation \\
\midrule
\multicolumn{5}{l}{\emph{qubits}}\\
QCNN~\cite{cong2019qcnn} & none & quantum states & phase recognition & \\
circuit-centric~\cite{schuld2020circuit} & none & small benchmarks & simulation & \\
quantum kernels~\cite{havlicek2019feature} & none & few features & superconducting chip & \\
transfer learning~\cite{mari2020transfer} & pretrained CNN in front & images & IBM and Rigetti chips & \\
QGrad-CAM~\cite{lin2024qgradcam} & CNN in front & MNIST (10), dogs vs cats, TIMIT speech & simulation & gradient (Grad-CAM) \\
QCxAI~\cite{mehta2025qcxai} & none & $2\times2$ pixels & explanation study & parameter-shift saliency \\
gate Shapley~\cite{heese2025shapley} & none & small circuits & explanation study, also on chips & Shapley values of gates \\
xAI for QML~\cite{steinmuller2022xai} & none & $2\times2$ pixels & Rigetti chip & IG, SHAP on chip \\
HattriQ~\cite{dibrita2025hattriq} & none & MNIST, Fashion-MNIST, Bars and Stripes, NIST digits, transverse-field Ising phases & simulation, also noisy & IG (numerical) \\
\rowcolor{ours}this work, qubits & linear mask, off chip & colour Caltech-10, all pixels & 37.7\,\% on chip & exact; $2\times2$ on chip, $r = 0.88$ (0.82) \\
\midrule
\multicolumn{5}{l}{\emph{light}}\\
D$^2$NN~\cite{lin2018d2nn} & none & MNIST (10) & 91.75\,\% (sim.) & \\
D$^2$NN, improved training~\cite{mengu2019jstqe} & none & MNIST (10) & 97.18\,\% (sim.) & \\
differential D$^2$NN~\cite{li2019differential} & none & grey CIFAR-10 & 48.51\,\% (sim.) & \\
ensemble of 30~\cite{rahman2021ensemble} & weighted sum of scores & grey CIFAR-10 & 62.13\,\% (sim.) & \\
magneto-optical~\cite{fujita2022mod2nn,watanabe2026mo} & none & MNIST; colour CIFAR-10 & $>$90\,\%; 38.6\,\% (sim.) & \\
GPAM~\cite{huang2026gpam} & none & MNIST, Fashion-MNIST & 94.37\,\%; 92.75\,\% (sim.) & gradient maps \\
\rowcolor{ours}this work, light & none & colour Caltech-10, CIFAR-10, all pixels & 72.3\,\%; 47.9\,\% (sim.) & exact; per pixel, simulated bench \\
\bottomrule
\end{tabular}
\caption{This work beside earlier quantum and diffractive classifiers (a comparison of kind, not a ranking).
``Classical network?'' asks whether a trained classical network sits before or after the quantum or optical part. An
empty explanation cell means explanations were not the work's aim. \emph{All pixels}: every pixel and colour of the
$32\times32$ picture ($3\times1{,}024$ values) is an amplitude, with no PCA, pooling or feature extraction. Our qubit result: 300 test pictures on
ibm\_kingston; our optical results use 5,124 parameters; the differential D$^2$NN has $2\times10^5$ neurons; the
magneto-optical MNIST figure is as reported there.}
\label{tab:compare}
\end{table}

Two points of this setting do not show in the table alone. Colour is handled with one wavelength: the channels and
their means travel as tiles of one field rather than as separate wavelengths, and every pixel and colour of the
$32\times32$ picture is an amplitude. And the decomposition needs no approximation, while one identity estimates it from
measurements: on a quantum computer and, from camera frames, on a (simulated) lens bench. This sets them beside the
gradient~\cite{lin2024qgradcam}, Integrated-Gradients and Shapley~\cite{steinmuller2022xai,dibrita2025hattriq},
parameter-shift~\cite{mehta2025qcxai} and gate-Shapley~\cite{heese2025shapley} explanations proposed for quantum
models, and the gradient maps of diffractive networks~\cite{huang2026gpam}, which serve other models and questions.
For Integrated Gradients with a zero baseline, pixels that are exactly zero receive zero attribution by
construction, as on the black backgrounds of MNIST; we therefore tested our maps on colour photographs in which object and background both carry signal: against
human-drawn object outlines, by deletion, and by a shortcut in the data that the maps exposed.

What the setting does not do is match electronic networks or the largest diffractive models in accuracy. A small
convolutional network reaches 85.4\,\% on Caltech-10 where the plates reach 72.3\,\%, and ensembles of thirty
diffractive networks reach 62\,\% on grey CIFAR-10 where our single four-plate model reaches 37.6\,\% (47.9\,\% in
colour). A less constrained quadratic model helps separate the physical constraints from the rest: the same
encoding, readout and training with a free complex map in place of the mask and plates (row Q of
Table~\ref{tab:ablation}) reaches $75.7 \pm 1.8\,\%$. Like ours, it is a ratio of quadratic forms of the encoded
picture. The four plates come within 3.3 points of it with 0.5\,\% of its parameters, in this training budget; this
does not bound what quadratic models in general can reach. The aim was a classifier whose class probabilities decompose exactly into contributions of
its encoded input that can be measured on hardware; these numbers are its price today.

\section{Limitations}
\label{sec:limits}
\begin{itemize}[leftmargin=*]
\item \emph{Loading on qubits.} The chip receives each picture through a fitted circuit (state fidelity 0.77 for the
masked states of D$'$, 0.96 for the plain ones of A$'$); for D$'$ this, not the chip's noise, costs most of the hardware
accuracy (Figure~\ref{fig:ibmsteps}), for A$'$ the chip's noise does (Appendix~\ref{app:checks}). Exact loading of $2^{12}$ amplitudes
with the general scheme of~\cite{plesch2011stateprep} takes about $\tfrac{23}{24}\,2^{12} \approx 3{,}900$ CNOTs in leading
order, so direct amplitude-encoded image classification, without a trained classical feature extractor
or classifier, stays limited on today's chips by exact loading, and by the chip's noise once a short fitted loading
replaces it, the known caveat of quantum machine learning on classical data~\cite{aaronson2015fineprint}.
\item \emph{The mask is computed off the chip.} In the hardware runs the Fourier-plane mask acts on the picture before
the chip; on a bench this is the 4f system in front of the plates, and on qubits a QFT with post-selected ancillas that
we simulate but did not run.
\item \emph{Scope of the hardware results.} Two trained models (D$'$ and A$'$, seed 0), one run of the test set each; for
D$'$ only, $2\times2$ maps of 30
pictures drawn at random and two runs of ten chosen ones, and $4\times4$ maps of three pictures; finer maps need more
shots than our quota allowed (at 4,000 shots a $4\times4$ block is at the shot-noise limit). A map measured on today's chip is not yet shown to be a more sensitive check of the device than its
accuracy: in simulation, every deliberate fault we tried that moves the maps beyond the chip's error at 4,000 shots
also costs the noise-free model at least 9.7 points of accuracy (Appendix~\ref{app:checks}). On the other test pictures
the measurement was simulated with shot noise but without gate noise or the fitted loading. The phase-plate model and its map measurement were checked in a second simulator, not built; that simulator
shares the model's physics (scalar, paraxial, thin plates, ideal lenses, the four channels kept apart), so it checks
the code, not these idealisations. Not modelled: how a modulator writes signed and complex values and filters the
unwanted diffraction orders, light crossing between the four channels of the 4f array, non-paraxial angles at the
pixel edges, and the camera's bit depth, noise and full well, which decide how many frames a per-pixel map needs in
practice. The simulations support the
feasibility of an optical implementation, but do not establish its experimental accuracy or the fidelity of its
evidence maps. A physical bench must still demonstrate signed-field and means-plane preparation, alignment and
magnification within the reported tolerances, and stable or reference-normalised optical power across the frames of a
map.
\item \emph{What we do not claim.} The all-optical model is a linear diffractive network: classical light suffices and
no quantum advantage is claimed. Four masks in superposition gave the same accuracy as four masks measured apart (a
no-interference control; Appendix~\ref{app:bank}), so no interference advantage is claimed either. The datasets are small images of ten
classes; accuracies are below those of electronic networks.
\item \emph{The readout.} BHR replaces a learned temperature by a chosen constant $\varphi$; it is not more accurate
than a learned temperature or than a softmax at the same fixed slope, accuracy stays within the seed spread only for slopes from
about 68 to 272, and as trained its confidences are not calibrated (ECE 0.10). Its anchor $b_g$ is the group mass of
a maximally mixed input, exact for unitary circuits; on a lossy bench it is kept as a fixed reference
(Section~\ref{sec:bhr}).
\item \emph{Stability of pixel maps.} Each map explains its own model exactly, but models trained from different seeds
use different single pixels: the maps of two seeds agree little pixel by pixel and much more in blocks
(Table~\ref{tab:seeds}), and a model and a bench that agree to an overlap of 0.997--0.998 give per-pixel maps (means plane included, as the bench shows it) that correlate at
0.88--0.94 (13 pictures) (Section~\ref{sec:bench}). Single-pixel detail is therefore a property of one trained model; block sums and
the share on the object (Table~\ref{tab:object}; the share of evidence inside the outline varies by at most 0.01
between seeds, its ratio to the outline's area by at most 0.03) are the robust readings.
Maps of the plates and of the qubit circuit agree about as much as those of two qubit seeds do (on pictures both
classify correctly 0.15 against 0.16), but less in the coarsest blocks (0.42 against 0.56, last row of Table~\ref{tab:seeds}): at the scale of pixels and small
blocks the two systems differ in their maps no more than the less stable of them differs from itself, in the coarsest
blocks somewhat more.
\item \emph{Test-set reuse.} The trim that removed the shortcut was chosen after the shortcut was seen in maps of test
pictures (Section~\ref{sec:shortcut}). The trimmed Caltech-10 results are therefore exploratory: validation-only model
selection afterwards does not undo that earlier use, and no independent holdout confirms the choice; a retrospective
analysis of the training pictures supports the rationale for the trim.
\item \emph{Dataset bias.} Trimming removed one shortcut; others belong to the data (motorbikes photographed on white
studio backgrounds, ketches on water) and remain visible in the maps.
\end{itemize}
\begin{table}[!htb]
\centering\footnotesize
\setlength{\tabcolsep}{2.8pt}
\begin{tabular}{lccccc}
\toprule
& \multicolumn{4}{c}{all 300 pictures} & both right \\
\cmidrule(lr){2-5}\cmidrule(lr){6-6}
model & pixels & blocks of 4\,px & blocks of 8\,px & blocks of 16\,px & pixels \\
\midrule
plates (row D)  & 0.25 (0.16--0.35) & 0.69 (0.55--0.79) & 0.78 (0.63--0.88) & 0.89 (0.61--0.96) & 0.29 (0.19--0.40) \\
qubits (D$'$)   & 0.10 (0.02--0.18) & 0.26 (0.09--0.43) & 0.40 (0.16--0.63) & 0.56 ($-$0.09--0.87) & 0.16 (0.09--0.27) \\
\midrule
plates vs qubits & 0.09 (0.02--0.19) & 0.28 (0.07--0.48) & 0.37 (0.08--0.64) & 0.42 ($-$0.16--0.82) & 0.15 (0.06--0.26) \\
\bottomrule
\end{tabular}
\caption{Agreement of the evidence maps between models trained with seeds 0, 1 and 2: per-picture correlation of the
true class's evidence for each pair of seeds, median (quartiles), 300 test pictures; blocks sum the pixel map. The last
row pairs each plate seed with each qubit seed (9 pairs). The last column keeps the pictures that both models of a
pair classify correctly (587, 373 and 1,316 picture pairs).}
\label{tab:seeds}
\end{table}

\section{Conclusion}
In a classifier that is linear in its encoded amplitudes up to a Born-rule measurement, of light or of qubits, each
class's probability above its share splits exactly into contributions of the encoded amplitudes, shown as Born
evidence maps; with the transmission held fixed, the split does not depend on the readout that follows. Interpretability here is built into the
architecture rather than estimated after training. These maps can be measured on the quantum chip that runs the classifier and, in simulation, on a lens bench, and in our tests they were useful: in a deletion test they ranked pixels as well as occlusion and gradient $\times$ input, they showed that
the mean-centred encoding moves the input light, and with it the evidence, onto the object, and they found a
shortcut in the data. On a superconducting quantum computer, with the mask applied before loading, the whole test set ran at 3.8 times
chance, and the explanations measured for 30 test pictures drawn at random correlated with the exact ones
($r = 0.88$; 0.82 with the standard estimator); for this masked model loading the picture, not the chip, costs the
most, while a model with no mask lost as much to the chip's noise instead. As the trim of Caltech-10 was chosen on test pictures, the Caltech-10 results are exploratory. The phase-plate version needs no quantum hardware at all; in simulation its main demand on a builder is
sideways alignment of the plates to about a micron. A single learned Fourier-plane filter already added
about four points (rows B to D of Table~\ref{tab:ablation}); several filters, side by side as channels or in series with
the plates, would keep the model linear up to the measurement and its explanations exact, and are a natural next
step. The code, the trained models, the data and a ledger of the hardware jobs are released
with the paper, with notebooks to try the maps and to train a model in the simulator.

\section*{Acknowledgements}
We acknowledge the use of IBM Quantum services for this work. The views expressed are those of the authors, and do
not reflect the official policy or position of IBM or the IBM Quantum team.

\section*{Code and data availability}
The code, every trained Caltech-10 model (all seeds), the scripts of the Caltech-10 analyses (deletion tests,
baselines, the readout sweep, model randomisation, bench tolerances, the LightPipes bench, the checks of the
hardware maps) with their outputs, Caltech-10 as
prepared here (derived from Caltech-101, CC BY 4.0), the record of every job of the test-set and map runs on ibm\_kingston (circuits as sent,
calibration and raw counts), and two notebooks (an interactive demo and training in the simulator) are at \url{https://github.com/stein-research/seeing-why}; release v1.2, which this version of the paper reports, is archived at Zenodo,
\href{https://doi.org/10.5281/zenodo.23984169}{doi:10.5281/zenodo.23984169}.
The computing setup and run times are in Appendix~\ref{app:compute}.

\section*{Author contributions and use of AI tools}
T.T. conceived the study, developed the methods and the code, ran the simulations and the hardware experiments, and
wrote the manuscript. R.J. worked with T.T. on the physics and the optics of the model, reviewed the manuscript and
advised on its content and presentation. An AI assistant (Claude, Anthropic) was used under the authors' direction to
write and run simulation and analysis code, to prepare and check the hardware jobs, to write the code that draws the
figures, to search the literature and check the references against their sources, and to draft and revise the text;
the authors chose what to cite. The authors retained responsibility for the study design, the scientific judgments
and the claims, checked the reported results against the saved data and the code, approved the final version, and
take full responsibility for its content.

\appendix
\section{Model and training details}
\label{app:details}
\emph{Mask parameters.} Per colour plane $c$ and spatial frequency $k$ the mask is
$d_c(k) = \cos\theta_c \cos\bar\theta_k\, e^{i\bar\phi_k}$, with $\bar\theta_k = (\theta_k + \theta_{-k})/2$ even and
$\bar\phi_k = (\phi_k - \phi_{-k})/2$ odd in $k$, so the filter's kernel is real. The amplitude adds 518 parameters
(row C against B in Table~\ref{tab:ablation}: 514 frequency angles and one gain for each of the four planes) and the phase 510 more (row D; the zero and Nyquist frequencies, their
own mirrors, have no phase). With the lens mask every frequency of the picture zero-padded
to $64\times64$ takes the mask value of the modulator pixel it falls on.
\emph{Qubit circuit.} Rotations $R_X(w_1)R_Y(w_0)R_Z(w_2)$ per wire and layer; CNOT$(i, i{+}1)$ for $i \equiv k \pmod 2$
after layer $k$; qubits ordered most significant first; the measured qubits are the first seven.
\emph{Phase plates.} Plate pixels of $2\times2$ field samples ($32\times32$ plate pixels on the $64\times64$ field);
transfer function on a grid padded to $128\times128$, cropped back after every step (light outside the aperture is
lost); detector regions $4\times8$ field pixels. The pixel-aware variant uses $2\times2$ samples per field pixel and
$z$ scaled by 4; the adjoint sums the samples.

\emph{Quadratic baseline (row Q of Table~\ref{tab:ablation}).} A free complex matrix $W$ ($128\times4096$, real and
imaginary parts trained, entries drawn from a complex normal distribution of variance $1/4096$, no regulariser or
constraint) maps the encoded picture to 128 outcomes, $P_b = |(Wa)_b|^2/\|Wa\|^2$, grouped as for the plates and read
by the same BHR ($\varphi = 0.0556$). Training as for row D: 50 epochs, class-balanced batches of 64, random crops
0.8, Adam 0.01 with a one-epoch warm-up and cosine decay, the weights of the best epoch on the 100 validation pictures
kept, seeds 0--2. It reaches 0.97--0.99 on training pictures and 0.87--0.89 on validation. Unlike the plates, $W$ is
neither a product of free-space propagation and phase masks nor limited to the losses of an aperture. Code:
\texttt{quad\_baseline.py}.

\section{One picture through the trained models}
\label{app:inside}
\begin{figure}[!htbp]
\centering
\includegraphics[width=0.72\linewidth]{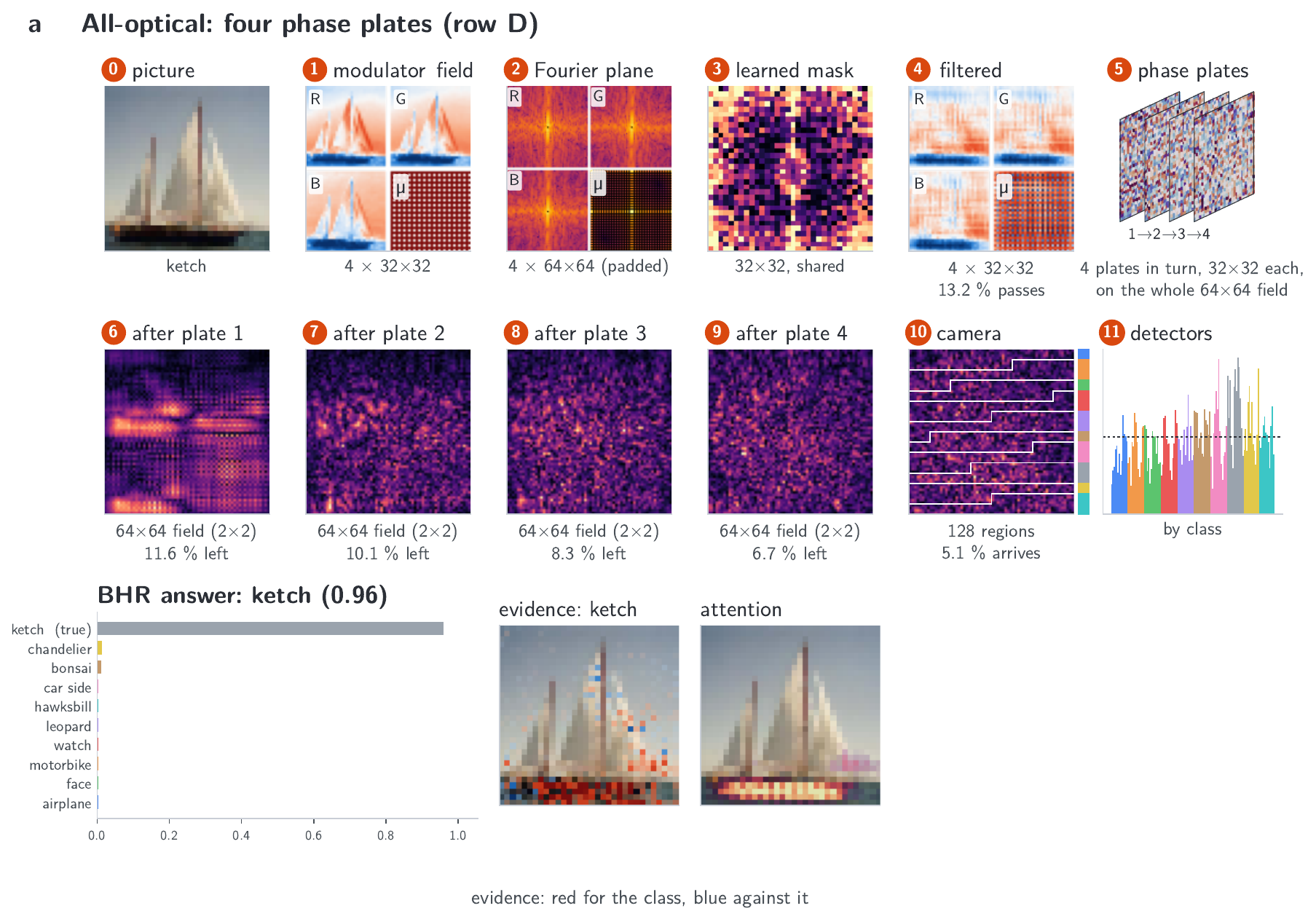}\\[1mm]
\includegraphics[width=0.72\linewidth]{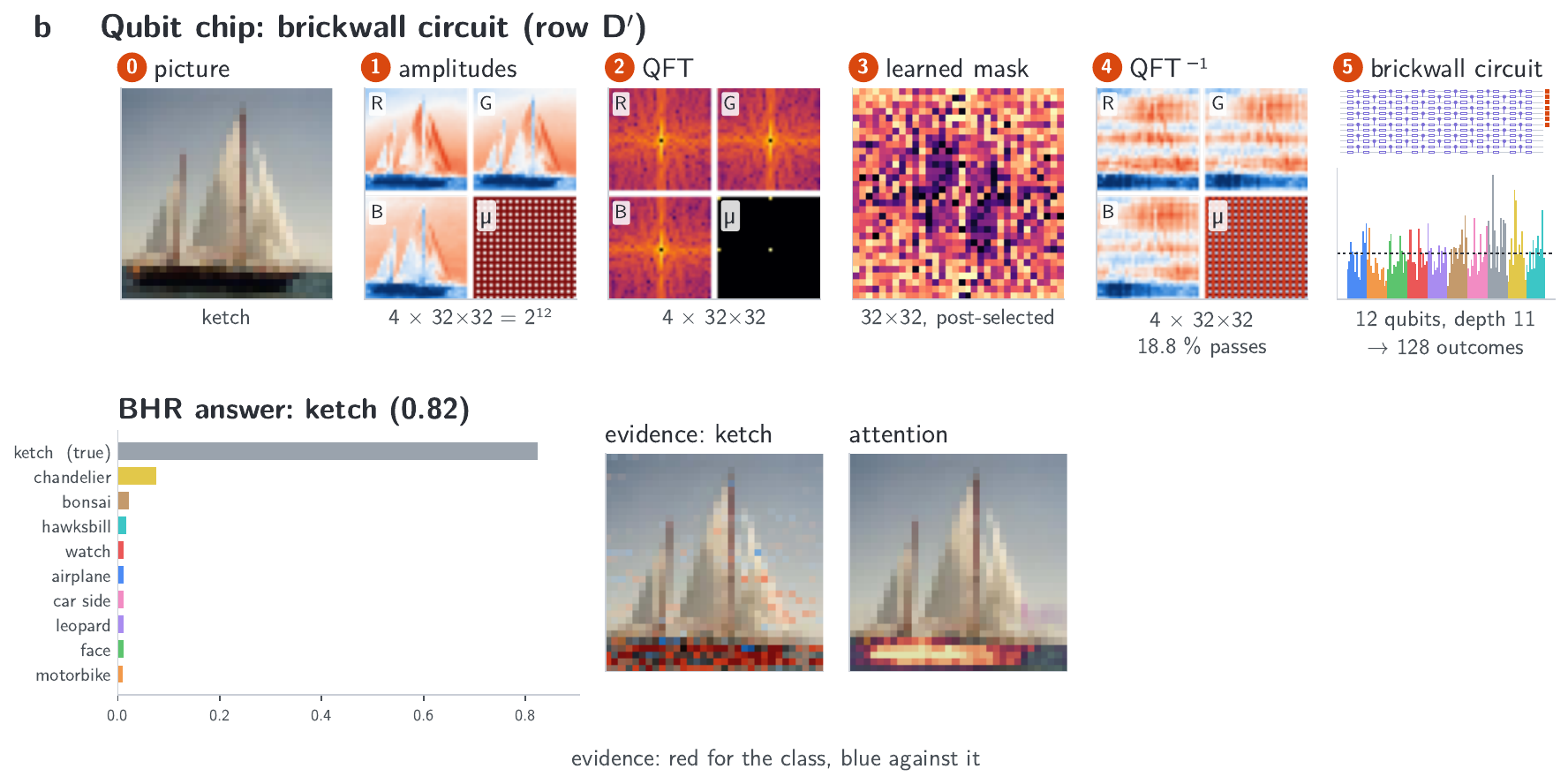}
\caption{The ketch of Figure~\ref{fig:overview} through the trained models: (a) the phase plates of row D, (b) the
qubit circuit of row D$'$. Colours as in Figure~\ref{fig:overview}: the modulator field (amplitudes) and the filtered field's
real part in red for positive and blue for negative, white zero, each plane on its own scale; the mask by its
magnitude; the plates' phase on a cyclic scale; after each plate, the field's magnitude. The four planes are tiled
$2\times2$ (R, G, B and the means plane). The plates are numbered 1--4. \emph{Passes}: the share of the light
through the Fourier-plane mask (on qubits, the post-selection success). \emph{Left}: the share of the input light still
in the aperture after each plate. \emph{Arrives}: the share that reaches the camera regions. Camera: square root of the
intensity; white lines separate the regions of different classes, the strip gives each row's class. Detectors and
outcomes: the light of the 128 regions (on qubits, the 128 measured bit strings), coloured by class; dashed: the uniform
share $1/128$. On qubits, step 5 sketches the brickwall circuit (boxes: rotations; links: CNOTs; squares: the seven
measured qubits) above its 128 outcomes. Bottom row of each panel: the BHR class probabilities, the exact evidence for the true class over the
picture (red for the class, blue against it) and the attention (each pixel's share of the light that reaches the measurement;
its positive part, over the picture).}
\label{fig:inside}
\end{figure}

Figure~\ref{fig:inside} follows the ketch of Figure~\ref{fig:overview} through the trained models of rows D and
D$'$, step by step, with the numbers each step produces. The interactive demo (Code and data availability) draws the
same view for any test picture and model. The four plates are drawn side by side but act one after another, each over
the whole tiled field of all four planes; each has $32\times 32$ pixels, and each pixel covers $2\times 2$ samples of the
$64\times 64$ field.

\section{Depth}
\label{app:depth}
On the untrimmed crops (which carry the shortcut of Section~\ref{sec:shortcut}, so these numbers are a record, one
seed): the brickwall at depths 22 / 11 / 6 reaches 58.7\,\% / 56.3\,\% / 45.0\,\% plain (ROC-AUC 0.884 / 0.868 / 0.802) and
63.7\,\% / 60.0\,\% / 53.7\,\% with the mask and the mean-centred encoding (0.918 / 0.893 / 0.863). With the mask and the mean-centred encoding, half the depth exceeds the plain circuit (depth 11 at 60.0\,\% against
depth 22 at 58.7\,\%), and depth 6 gains nine points; the mask alone does not help the shallow circuits (depth 6:
44.0\,\% with it, 45.0\,\% without).

\section{Four masks and the interference control}
\label{app:bank}
Four Fourier-plane masks in superposition, selected by two register qubits and mixed by the main circuit, reach
$69.1 \pm 2.1\,\%$ on the untrimmed crops (three seeds; one mask $59.1 \pm 2.9\,\%$). A control with the same circuit and
parameters, in which the register is measured after the masks so that the filtered pictures never interfere, reaches
69.3\,\% against 70.3\,\% at the same seed, with the same ROC-AUC (0.938). The gain comes from having four filters, not from
their interference, and we claim no interference advantage.

\section{Hardware details}
\label{app:hardware}
\begin{figure}[!htb]
\centering
\includegraphics[width=\linewidth]{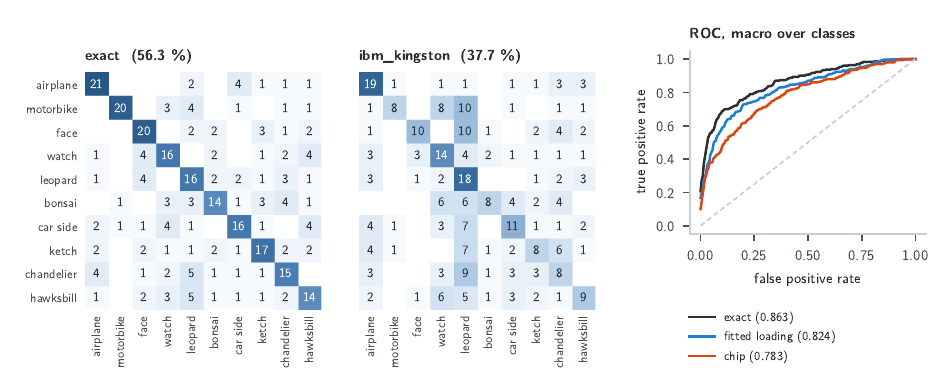}
\caption{The 300 test pictures: confusion matrices of the exact simulation and of ibm\_kingston, and ROC curves (macro over
classes) for the exact model, the fitted loading and the chip.}
\label{fig:hwcm}
\end{figure}

\emph{Loading.} For one test picture (with an earlier model trained on the untrimmed crops) the fitted RY--CNOT preparation of depth 8 (fidelity 0.75, 44 CNOTs; 105 with the trained circuit) gave
the true class a probability of 0.950 under a noisy simulation (FakeFez, the backend model available then; every later simulation uses FakeKingston) and of 0.771 on ibm\_kingston, against 0.626 and 0.686
for depths 16 and 24 in simulation: deeper loading is more exact but noisier, and depth 8 was kept. For the test set, all 300
preparations were fitted at once (3,000 Adam steps, 10 minutes on one GPU). Each circuit was transpiled at Qiskit's
level 3 with eight seeds and the one with the best calibrated success estimate kept (the brickwall always maps onto a line
of twelve qubits of the heavy-hex lattice without SWAPs; every test-set circuit of D$'$ used the same line, and 295 of
the 300 of A$'$ the same twelve qubits in the reverse order, so A$'$ read its seven outcome qubits on other qubits; the map
jobs used other lines); dynamical decoupling (XY4) was on. The bit order
of the counts was checked end to end against our simulator (total variation distance 0.011 at 200,000 shots, logical
and transpiled; a reversed order would give 0.24).
\emph{Pictures and jobs.} Table~\ref{tab:jobs} lists every job of the test-set and map runs; the one-picture loading test above (job \texttt{dbinnaune4hc73bf0tg0}) and a first one-picture map, both sent before the ledger was kept, are not in it. Pictures were sent one of each class in turn, ten jobs of thirty; per job the chip's
accuracy for D$'$ (23.3\,\%--50.0\,\%) followed the fitted loading's on the same pictures (26.7\,\%--53.3\,\%). Every job of the table is kept in a
ledger: job id, backend, times, quantum seconds, the calibration at sending, the qubits used, the circuits as built
and as sent, and the raw counts; a measurement is never sent twice by accident (keys over the circuits, the backend and the
shots), and deliberate repeats are marked.
\begin{table}[!htb]
\centering\small
\setlength{\tabcolsep}{4pt}
\begin{tabular}{L{0.56\linewidth}rrr}
\toprule
run & jobs & circuits & quantum s \\
\midrule
test set, model D$'$ (2026-10-31) & 10 & 300 & 340 \\*
\multicolumn{4}{L{0.94\linewidth}}{\scriptsize\ttfamily dbit0pohoj4c73b6str0 dbit2c6ne4hc73bf93b0 dbit2rmne4hc73bf9490 dbit3cohoj4c73b6t2og dbit3s6ne4hc73bf96m0 dbit4c8hoj4c73b6t4m0 dbit4tune4hc73bf98o0 dbit5d8hoj4c73b6t6pg dbit5t6ne4hc73bf9aug dbit6e5ep8hs73aohpsg} \\[2pt]
earlier maps, first run (2026-10-31) & 1 & 80 & 87 \\*
\multicolumn{4}{L{0.94\linewidth}}{\scriptsize\ttfamily dbitsbghoj4c73b6u5s0} \\[2pt]
earlier maps, second run (2026-10-31) & 1 & 90 & 97 \\*
\multicolumn{4}{L{0.94\linewidth}}{\scriptsize\ttfamily dbitvl01h92s738469gg} \\[2pt]
$2\times2$ maps, 30 random pictures (2026-11-03) & 6 & 390 & 426 \\*
\multicolumn{4}{L{0.94\linewidth}}{\scriptsize\ttfamily dbl425tgg49c73dk0di0 dbl43u5gg49c73dk0h60 dbl44oonceqs73an02c0 dbl45ll6p51s73cno1ag dbl46g8nceqs73an04cg dbl47bt6p51s73cno350} \\[2pt]
$4\times4$ maps, three pictures (2026-11-03) & 3 & 99 & 111 \\*
\multicolumn{4}{L{0.94\linewidth}}{\scriptsize\ttfamily dbl487d6p51s73cno42g dbl48oonceqs73an06rg dbl49bhkdqps7393i460} \\[2pt]
test set, model A$'$ (2026-11-04) & 10 & 300 & 340 \\*
\multicolumn{4}{L{0.94\linewidth}}{\scriptsize\ttfamily dbljva58mdtc73d69s70 dbljvq9kbkrc73cnrs4g dblk0a1kbkrc73cnrtj0 dblk0ou74qds73dqkdg0 dblk17m74qds73dqkeng dblk1me74qds73dqkfj0 dblk25l8mdtc73d6a3q0 dblk2le74qds73dqkhd0 dblk34l8ft8s738t7b50 dblk3km74qds73dqkj2g} \\[2pt]
\midrule
all & 31 & 1,259 & 1,401 \\
\bottomrule
\end{tabular}
\caption{Jobs on ibm\_kingston (Heron r2), with their IBM job ids, Qiskit 2.6.1 and qiskit-ibm-runtime 0.51.0:
SamplerV2, 4,000 shots, optimisation level 3, the best of eight transpiler seeds, dynamical decoupling (XY4). A fourth
$4\times4$ job was prepared and not sent. Each job's circuits as built and as sent, calibration and raw counts are in
the released ledger.}
\label{tab:jobs}
\end{table}
\emph{Maps.} The ten pictures of Figure~\ref{fig:hwmaps} were the most cluttered of those the chip classified correctly
with probability at least 0.6 (clutter: background share of the crop times the grey-level spread outside the object
outline; motorbike, which had none, took its most confident correct picture). Block preparations were fitted like the pictures (fidelity median 0.81). The 30 random pictures were
drawn with \texttt{numpy.random.default\_rng(20261103)}, three per class in class order, and the record was committed
before any fit; their $2\times2$ block preparations have fidelity median 0.78 (minimum 0.58; with the $4\times4$ ones, 0.77 and 0.52). Their whole-picture circuits, run
before in the test set, were sent again in the map jobs, marked in the ledger as repeats. A fourth $4\times4$ job,
prepared, was not sent for lack of quota. On the chip the errors of D$'$ drift towards leopard and watch (Figure~\ref{fig:hwcm}): it
answers leopard for 77 of the 300 test pictures and watch for 43, against 30 of each.

\Needspace{0.45\textheight}
\section{Computational setup}
\label{app:compute}
Everything except the hardware runs was done on one laptop (Table~\ref{tab:compute}).
\begin{table}[!htb]
\centering\small
\begin{tabular}{p{0.40\linewidth}p{0.54\linewidth}}
\toprule
\multicolumn{2}{l}{\emph{hardware}} \\
GPU, CPU, memory & NVIDIA GeForce RTX 5050 Laptop GPU (8\,GB); Intel Core i5-13450HX; 16\,GB \\
quantum computer & IBM ibm\_kingston (Heron r2) through Qiskit Runtime \\
\midrule
\multicolumn{2}{l}{\emph{software}} \\
Python 3.12.3 & \\
PyTorch 2.11 (CUDA 12.8)~\cite{paszke2019pytorch} & simulator, optics, training \\
NumPy 2.5, SciPy 1.18 & arrays, data \\
scikit-learn 1.9.1~\cite{pedregosa2011sklearn} & ROC-AUC, logistic-regression reference \\
OpenCV 5.0 & random crops \\
PennyLane 0.45.1~\cite{bergholm2018pennylane} & cross-check of the simulator \\
Qiskit 2.6.1~\cite{javadi2024qiskit} & cross-check, circuits, transpiling \\
qiskit-ibm-runtime 0.51.0 & hardware jobs \\
qiskit-aer 0.18.2 & noisy simulations (FakeKingston) \\
LightPipes 2.1.5~\cite{lightpipes} & independent optics simulation (the bench) \\
umap-learn 0.5.12~\cite{mcinnes2018umap} & Figure~\ref{fig:umap} \\
\midrule
\multicolumn{2}{l}{\emph{time}} \\
training, 50 epochs & 23--48\,s phase plates (row D; by seed); 2.7--6.7\,min qubit circuit (D$'$, 12-qubit state vector; by seed) \\
loading circuits & 300 fitted in about 10\,min \\
ibm\_kingston & 340 quantum seconds (test set, D$'$), 340 (test set, A$'$), 184 (two runs of the earlier maps), 537 (maps of the 30 random pictures: 426 for $2\times2$, 111 for $4\times4$) \\
\bottomrule
\end{tabular}
\caption{Computational setup.}
\label{tab:compute}
\end{table}

\section{Bench simulation}
\label{app:bench}
LightPipes (BSD-3), 633\,nm, 8\,\textmu m pixels: each colour plane through its own 4f system (focal length chosen so
the Fourier plane falls on the simulation grid, $4\times$ sampling per pixel, grid padded $2\times$), the learned mask
on a modulator covering $|k| \le 16$ and divided by the square pixel's sinc envelope; the 4f output passed to the
plates unaveraged; the planes tiled; Fresnel
propagation (\texttt{Forvard}) over $d = z\,\Delta x^2/\lambda = 3.24$\,mm between plates with square pixels; light
summed over each detector region. Propagation alone agrees with ours to $2\times10^{-10}$; with no mask the 4f system
images the picture to $7\times10^{-7}$. For the thick-plate check, each plate is a fused-silica relief ($n = 1.457$) of height
$h(x) = [\phi(x) \bmod 2\pi]\,\lambda/[2\pi(n-1)]$, at most 1.385\,\textmu m, crossed in 32 layers of LightPipes'
finite-difference \texttt{Steps} (version 2.1.5). On a uniform slab of thickness $t = \lambda/[4(n-1)]$, whose
expected phase is $k(n-1)t = \pi/2$, \texttt{Steps} returns $\pi/4$ (0.500 of it, measured at the start of every run);
the plates are therefore given the index $1 + (n-1)/0.500 = 1.914$, which restores the phase $k(n-1)h$ and is not
fitted to our model. With it, a single thick plate matches its thin phase mask to an overlap of 0.998. (\texttt{Steps}
also takes its index array transposed, which the code accounts for.) The check is consistency under this corrected
software convention, not an independent model of the glass. Code: \texttt{optics\_check\_full.py}, option
\texttt{--thick}. With 8\,\textmu m modulator pixels and one mask pixel per spatial frequency, a
real lens pair has $f = s\,\Delta x\,p/\lambda \approx 3.2$\,mm ($s = 32$ pixels of $\Delta x = 8$\,\textmu m, mask
pitch $p = 8$\,\textmu m); a longer focal length spreads each frequency over several mask pixels. Dividing out the
sinc envelope asks for gains up to 2.1 (rows D and E), highest at the corners of the band, where the envelope is
lowest; a passive mask is scaled to a peak of 1 and then keeps 23\,\% of the light. The readout normalises the camera light,
so the predictions do not change, but the photon budget falls by that factor.

\bibliographystyle{quantum}
\section{Further checks}
\label{app:checks}
The checks below answer questions that the main text points to; Table~\ref{tab:checks} lists them with their
results, in the order of the paragraphs that follow.

\emph{Readout slope and calibration.} On row D (Section~\ref{sec:ablation}), a sweep of $\varphi$ (three seeds each) shows that only the slope matters: from $\varphi = 0.0278$
to $0.111$ (slopes 272 to 68) BHR gives $69.8$--$72.3\,\%$, within the spread of the seeds; slope 34 ($\varphi =
0.222$) loses 14 points and slope 544 four; and a softmax of the group sums with its slope fixed at BHR's 136 gives
$70.9 \pm 2.3\,\%$ (ROC-AUC $0.950 \pm 0.003$). Nothing in BHR beyond its slope matters for accuracy; what it adds is
that the slope is set before training, from $\varphi$, in units of headroom. The slope does set the confidences. On the
300 test pictures BHR at $\varphi = 0.0556$ is underconfident (mean top-class confidence 0.64 at 72.3\,\% accuracy;
expected calibration error, ECE, $0.10 \pm 0.02$ over the seeds, 15 bins), the learned temperature (slope about 380)
is overconfident (0.82 at 73.8\,\%; ECE 0.09), and the softmax at slope 136 has ECE 0.09. Temperature
scaling~\cite{guo2017calibration} of BHR, fitted on half of the test pictures and applied to the other half, lowers
its ECE to 0.06; of the sweep, $\varphi = 0.0278$ (slope 272) is the best calibrated as trained (0.07). BHR fixes the
slope for training, not for calibration.

\begin{table}[!htb]
\centering\small
\setlength{\tabcolsep}{4pt}
\rowcolors{2}{white}{stripe}
\begin{tabular}{L{0.24\linewidth}L{0.33\linewidth}L{0.36\linewidth}}
\toprule
check & question & result \\
\midrule
readout slope & does BHR's accuracy rest on its slope alone? & 69.8--72.3\,\% for slopes 68--272 (row D) \\
calibration & are the readouts' confidences calibrated? & BHR underconfident (ECE 0.10), 0.06 after temperature scaling \\
retraining on deleted pictures & can ROAR rank the maps? & no: the deleted pixels' pattern alone gives the class (90--98\,\%) \\
reach of the mask's kernel & how far does a block of the filtered state stand for the picture? & 66\,\% of the kernel's energy on the pixel itself \\
steps of the map error & where does the chip's map error come from? & preparation slope 1.09, noise 0.90, chip 0.82 \\
$4\times4$ maps & finer maps on the chip & $r = 0.60$, at the shot-noise limit \\
the earlier maps' first run & a term reused from another job & 0.63, lowered by selection on a noisy measurement \\
a deliberate fault & does a map see a fault that the accuracy does not? & not in our set: 81 of 246 faults seen, each costs $\geq 9.7$ points \\
model A$'$ on the chip & where does a model with no mask lose? & not to loading: the chip costs 22 points; on 100 pictures the noise model predicts half \\
\bottomrule
\end{tabular}
\caption{The further checks of this appendix, with their results.}
\label{tab:checks}
\end{table}

\emph{Retraining on deleted pictures.} Retraining on deleted pictures (ROAR~\cite{hooker2019benchmark}), the remedy for
deleted pictures leaving the training data (Section~\ref{sec:deletion}), does not help here.
We trained row D again from scratch on pictures with 10, 30, 50 and 70\,\% of their pixels deleted (three seeds
each) and tested it on test pictures deleted the same way: ranked by true-class evidence, 66.2, 68.9, 70.2 and
70.0\,\%; by intensity, 65.7, 59.2, 55.2 and 48.1\,\%; at random, 71.7, 67.2, 66.9 and 65.6\,\% (72.3\,\% with
nothing deleted). The evidence ranking chooses the pixels with the label, and the pattern of deleted pixels alone, with
no picture, gives the class to a logistic regression for 90--98\,\% of the test pictures (intensity 28--35\,\%,
random 10--11\,\%): the retrained model can read the label from where the pixels were taken, so ROAR cannot rank a
class-specific map.

\emph{Reach of the mask's kernel.} The chip's maps (Section~\ref{sec:hardware}) are maps of the filtered state, which
stands for the picture only as far as the mask's kernel reaches: the kernel of model D$'$ keeps two thirds of its
energy on the pixel itself (66\,\% at distance 0, 76\,\% within 8 pixels), and a median 67\,\% of a filtered
quadrant's energy comes from the same quadrant of the picture (74--80\,\% for seeds 1 and 2).

\emph{Steps of the map error.} On the 30 random pictures (Section~\ref{sec:hardware}), the fitted preparations alone inflate the maps (slope 1.09,
$r = 0.94$ without noise), the simulated noise then shrinks them (0.90), and the chip gives 0.82. Against the noisy
device's own maps, those of the noisy simulation of the same circuits without shot noise, the chip's maps correlate at
0.93 (centred 0.96, slope 0.91), so what remains is mostly shot noise.

\emph{$4\times4$ maps.} Each took 33 circuits (Eq.~\eqref{eq:flip}) in one job; the quota covered three of the four
pictures. They correlate with the exact maps at 0.60 (within-picture permutation, 95th percentile 0.19), with 73\,\%
same sign where 60\,\% of the exact blocks are positive, but the strongest block is found in none of the three, the
relative error is 1.6 and the blocks miss the job's own $P_g - b_g$ by 0.027 on average, more than ten times a block's
mean $|E_S|$ (0.0017). At 4,000 shots a $4\times4$ block is at the shot-noise limit, as the simulation predicted
($r = 0.68$ for the four pictures).

\emph{The earlier maps' first run.} A first run of the earlier maps (Figure~\ref{fig:hwmaps}), which reused each picture's whole-picture term $\bra{\psi}M_g\ket{\psi}$
from the test-set job an hour earlier, gave 0.626 (centred 0.86). That job's measurement had also chosen the pictures:
nine of them selected on a noisy $\hat p_g \geq 0.6$, their measured $P_g - b_g$ sit just above the threshold (0.022--0.032) while
the exact values spread from 0.005 to 0.043, the mark of selection on a noisy measurement. The term enters all four
blocks of a map, so it lowers the pooled correlation and leaves the centred one; the random pictures measure it in the
same job. Against the earlier maps, a null that spreads $P_g - b_g$ over the blocks in proportion to their light
$\|\psi_S\|^2$ correlates at 0.55 (centred 0.38), and calling every block positive matches 82.5\,\% of their signs
against the measured 88\,\%.

\emph{A deliberate fault.} Would a measured map show that the device does not run the circuit it is meant to? We
simulated, on ten of the random pictures (the first of each class), the trained circuit with one deliberate fault, as
if mis-calibrated: one rotation (RY or RZ) of a whole layer (layer 0, 5 or 11) off by 0.1 to 1.0\,rad, or one rotation
(RY or RX) of one gate in layers 0--2 off by 0.5 to 1.5\,rad, 246 faults in all. Of these, 81 move the noise-free
$2\times2$ maps by more than the chip's error on the random pictures' maps (mean absolute error 0.0025 per block,
sign flip, 4,000 shots), and each of them also costs the noise-free model at least 9.7 points of test accuracy (the cheapest: one RX gate of layer 1 off by 1.5\,rad, maps moved by 0.0032 per
block; for example, all RY angles of layer 0 off by 0.3\,rad move the maps by 0.0039 per block and the accuracy from
56.3\,\% to 31.3\,\%); the faults that cost at most five points move the maps by 0.0022 per block or less. In this set, a fault
large enough to rise above today's map error at this shot count is one the accuracy already shows, so we did not
spend chip time on the test; a fair test on the chip would compare the maps and the accuracy of the faulty and the
nominal circuit at equal shots.

\emph{Model A$'$ on the chip.} Model A$'$ (Table~\ref{tab:ablation}), the qubit circuit on plain amplitudes with
no mask (seed 0), ran on ibm\_kingston as D$'$ did (Section~\ref{sec:hardware}), four days later: the same 300 test
pictures, a fitted loading of depth 8, ten jobs of thirty, 4,000 shots, 340 quantum seconds, on the same twelve qubits
mostly in the reverse order (Appendix~\ref{app:hardware}; the readout errors of its outcome qubits at sending,
0.004--0.010, were no worse than those of D$'$). Its plain states load better (fidelity median
0.96, against 0.77 for the masked states of D$'$): 51.3\,\% exact and 54.7\,\% with the fitted loading ($p = 0.06$),
but the chip then costs it 22 points, to 32.3\,\% ($p < 10^{-12}$; ROC-AUC $0.855 \to 0.844 \to 0.758$). On the chip
the two models end close, 37.7 against 32.3\,\% (paired over the same pictures, $p = 0.16$; 64 right only with D$'$,
48 only with A$'$). The circuits are the same size as sent (105 two-qubit gates, depth 97--101). The backend's noise model
predicts part of the difference: on the first 100 pictures (ten per class), transpiled as sent, the noisy simulation
(FakeKingston, 4,000 shots) takes A$'$ from 57\,\% with the fitted loading to 47\,\%, and the chip to 37\,\%, while
D$'$ goes from 41 to 42\,\% and 38\,\%. The plain states are the more sensitive to the device's noise, in the
simulation as on the chip; we have not found why (the two models' margins between the top two classes with the fitted
loading are alike on these 100 pictures, median 0.20 and 0.23). The chip's per-job accuracy for A$'$ was 16.7\,\%--46.7\,\% (fitted loading
46.7\,\%--66.7\,\%), and its errors drift towards airplane, the answer for 70 of the 300 pictures.

\bibliography{refs}
\end{document}